\pdfoutput=1
\documentclass[twoside,leqno,onecolumn]{article}
\usepackage{ltexpprt} 

\usepackage[dvipdfmx]{graphicx}
\usepackage{amsmath}
\usepackage{tabularx}

\usepackage{algorithm,algorithmicx,algpseudocode}
\usepackage{listings}
\usepackage{color}
\usepackage{xspace}
\usepackage{caption}
\DeclareCaptionType{copyrightbox} 
\usepackage{subcaption}
\usepackage{alltt}
\usepackage{balance}
\usepackage{hyperref}
\usepackage{paralist}
\usepackage{cite}
\usepackage[normalem]{ulem}
\usepackage{soul}

\newcommand{\eat}[1]{}

\definecolor{gray}{rgb}{0.5,0.5,0.5}

\definecolor{light-gray}{gray}{0.15}

    {\begin{list}{}{
       \renewcommand{\makelabel}[1]{\bf $\bullet$}\hfil%
       \settowidth{\labelwidth}{\bf $\bullet$}%
       \setlength{\partopsep}{0mm}%
       \setlength{\parsep}{0mm}%
       \setlength{\parindent}{0mm}%
       \setlength{\itemsep}{0mm}%
       \setlength{\topsep}{0mm}%
       \setlength{\leftmargin}{\labelwidth}%
       \addtolength{\leftmargin}{\labelsep}
     }}%
    {\end{list}}

\newlength{\alglabelwidth}
\newcommand{\alginput}[1]{%
\par\noindent%
\settowidth{\alglabelwidth}{\emph{Output:}}%
\makebox[\alglabelwidth][l]{\emph{Input:}} \begin{tabular}[t]{l} #1 \end{tabular}}
\newcommand{\algoutput}[1]{%
\par\noindent%
\settowidth{\alglabelwidth}{\emph{Output:}}%
\makebox[\alglabelwidth][l]{\emph{Output:}} \begin{tabular}[t]{l} #1 \end{tabular}}

\title {Query-Driven Sampling for Collective Entity Resolution}

\eat{%
\author{\IEEEauthorblockN{Christan Grant}
    \IEEEauthorblockA{University of Florida\\
    Dept.\ of Computer Science\\
    Gainesville, Florida, USA\\
    cgrant@cise.ufl.edu}
\and
  \IEEEauthorblockN{Daisy Zhe Wang}
    \IEEEauthorblockA{University of Florida\\
    Dept.\ of Computer Science\\
    Gainesville, Florida, USA\\
    cgrant@cise.ufl.edu}
\and
  \IEEEauthorblockN{Michael Wick}
    \IEEEauthorblockA{University of Massachusetts, Amherst\\
    Dept.\ of Computer Science\\
    Amherst, Massachusetts, USA\\
    mwick@cs.umass.edu}}
}

\begin{document}
\maketitle

\begin{abstract}

Probabilistic databases play a preeminent role in the processing and management of uncertain data.
Recently, many database research efforts have integrated probabilistic models
into databases to support tasks such as information extraction and labeling.
Many of these efforts are based on batch oriented inference which inhibits a
realtime workflow. One important task is entity resolution (ER).
ER is the process of determining records (mentions) in a database that correspond to the same real-world entity.
Traditional pairwise ER methods can lead to inconsistencies and low
accuracy due to localized decisions. Leading ER systems solve this
problem by collectively resolving all records using a probabilistic
graphical model and Markov chain Monte Carlo (MCMC) inference.
However, for large datasets this is an extremely expensive process.
One key observation is that, such exhaustive ER process incurs a
huge up-front cost, which is wasteful in practice because most users
are interested in only a small subset of entities. 

In this chapter, we
advocate pay-as-you-go entity resolution by developing a number of
query-driven collective ER techniques. We introduce two classes of
SQL queries that involve ER operators --- {selection-driven}
ER and {join-driven} ER\@. We implement novel variations of the MCMC
Metropolis Hastings algorithm to generate biased
samples and selectivity-based scheduling algorithms to support the
two classes of ER queries. Finally, we show that query-driven ER
algorithms can converge and return results within minutes over a
database populated with the extraction from a newswire dataset
containing $71$ million mentions.
\end{abstract}

\section{Query-Driven Entity Resolution Introduction}

Entity resolution (ER) is the process of identifying and
linking/grouping different manifestations (e.g., mentions, noun
phrases, named entities) of the same real world object. It is a crucial task for
many applications including knowledge base construction, information extraction, and question answering. 
For decades, ER has been studied in both
database and natural language processing communities to link
database records or to perform entity resolution over extracted mentions (noun
phrases) in text.

ER is a notoriously difficult and expensive task.
Traditionally, entities are resolved using strict pairwise similarity, which usually leads
to inconsistencies and low accuracy due to localized, myopic decisions~\cite{Wick:2012:DHM:2390524.2390578}.
More recently, {collective} entity resolution methods have achieved
state-of-the-art accuracy because they leverage relational
information in the data to determine resolution jointly rather than
independently~\cite{Bhattacharya:2007:CER:1217299.1217304}. 
However, it is expensive to run collective ER based on probabilistic
graphical models (GMs), especially for {cross-document}
entity resolution, where ER must be performed over millions of mentions.

In many previous approaches, collective ER is performed
exhaustively over all the mentions in a data set, returning all entities.
Researchers have developed new methods to
perform large-scale cross-document entity resolution over parallel
frameworks~\cite{singh2011large,Wick:2012:DHM:2390524.2390578}.
However, {in many ER applications, users are only interested in one
or a small subset of entities.}
This key observation motivates
query-driven ER, an alternative approach to solving the
scalability problem for ER\@. 

Compared to previous ER models and algorithms, query-driven techniques
in this chapter scale to data sets that are in many cases three orders of magnitude larger.
Moreover, the ER model in this chapter is general enough to take both bibliographic 
records and mentions extracted from unstructured text.
Query-driven ER
techniques over GMs can also be generalized for other applications to
perform query-driven inference.

This work follows a line of research on implementing ML models inside of
databases~\cite{Wick:2010:SPD:1920841.1920942,hellerstein2012madlib,li2013gptext}.
Researchers use factor graphs because this flexible representation works well
with other machine learning algorithms.
ER is ubiquitous and an important part of many analytic pipelines; a
probabilistic database implementation is natural.

In this chapter, we first introduce SQL-like queries that involve ER
operations. These ER operators are an SQL comparison
operator (i.e., ER-based equality) that returns true if two mentions
map to the same entity. Factor Graphs, a type of GM, are used to
model the collective entity resolution over extracted mentions from
text. Using this ER based comparison operator, users can pose
selection queries to find all mentions that map to a single entity
or pose join queries to find mentions that map to the subset of
entities that they are interested in resolving. \eat{For web search
applications the query can be one entity, whereas for security
applications the query can be a watchlist of entities. <query
examples>}

\eat{Compared to the traditional exhaustive ER, where all mentions map to
a set of resolved entities as result, query-driven ER adopts a
different execution model. Intuitively, ER does not have to be
performed offline over the database/knowledge base. Given a query,
ER is performed on-the-fly. Query-driven ER algorithms focus
computation on the resolution of the mentions that are most-likely
to be mapped to the query entities, using biased MCMC sampling based
on approximate pairwise distance between mentions.}

Because exhaustive ER is expensive it is common to use {blocking} techniques
to partition the data set into approximately similar groups called canopies.
Query-driven ER in this chapter differs from blocking in two important ways:
1) deterministic blocks are replaced by a pairwise distance-based metric, and 
2) blocks (or canopies) are implicit to the query-driven ER data set and do not have
to be created in advanced.
The latter point, implicit blocking, is realized using a data structure created
based on the similarity to a query mention.
This data structure allows parameters to include or remove mentions from the working data set.
This property is similar to the iterative blocking technique~\cite{Whang:2009:ERI:1559845.1559870}, which
is shown to improve ER accuracy. Such an approach can dramatically
amortize the overall ER cost suitable for the pay-as-you-go paradigm
in dataspaces~\cite{dataspace}.

To support ER driven by queries, we develop three sampling algorithms for MCMC inference over graphical models.
More specifically, instead of a uniform sampling distribution, we sample on a distribution that is biased to the query.
We develop a query-driven sampling techniques that maximizes the resolution of the target query entity
 ({target-fixed}) and biases the samples based on the pairwise
similarity metric between mentions and query nodes 
 ({query-proportional}). 
We also introduce a {hybrid} method that performs query-proportional
sampling over a fixed target.
We develop two optimizations to the query-proportional and hybrid methods
to model the similarity and dissimilarity between the
mentions and the query entity, i.e., {attract} and {repel} scores.
In the first {target-fixed}
algorithm, we adapt the samples to resolve the query entity.
The second {query-proportional} algorithm, selects mentions
based on their probabilistic similarity to the query entity. The third
{hybrid} algorithm combines the two approaches.
A summary of approaches can be found in
Table~\ref{tab:algorithmsummarytable}.

\eat{During ER entities may contain a large amount of candidate mentions
that are very similar.
In the case, when the cluster of mentions is homogeneous,
we show a different sampling methods to compute and apply
influence function to generate a {repel} score for biased
sampling.}

When a user is interested in resolving more than one entity we
employ multi-node ER techniques.
To implement multi-node ER queries, single-node ER techniques may be naively performed iteratively to
resolve one entity at a time. 
However, such an
algorithm can lead to un-optimized resource allocation if the same
number of samples is generated for each target entity, or low
throughput if one of the entities has a disproportionately low convergence rate.
To alleviate this problem, we present 
three multi-query ER algorithms that schedule the
sample generation among query nodes in order to
improve overall convergence rate.

In summary, the contributions of this chapter are the following:
\begin{itemize}

\item We define a query-driven ER problem for cross-document, collective ER
over text extracted from unstructured data sets;

\item We develop three single-node algorithms that perform focused
sampling and reduce convergence time by
orders-of-magnitude compared to a non-query-driven baseline (Section~\ref{sec:algorithms}).
We develop two influence functions that use {attract} and {repel}
techniques to grow or shrink query entities (Section~\ref{sec:vose});

\item We develop scheduling algorithms to optimize the overall
convergence rate of the multi-query ER (Section~\ref{sec:queryproper}). The best scheduling
algorithm is based on selectivity of different target entities (Section~\ref{sec:coinfluence}).
\end{itemize}

The results show that query-driven ER algorithms is a promising
method of enabling realtime, ad-hoc, ER-based queries over large data sets.
Single node queries of different selectivity converge to a
high-quality entity within $1$-$2$ minutes over a newswire 
data set containing 71 million mentions. Experiments also show that
such real-time ER query answering allows users to iteratively refine ER queries
by adding context to achieve better accuracy (Section~\ref{sec:experiments}).



\eat{In the rest of this chapter, we first introduce background
knowledge in Section~\ref{sec:background}. A system overview is
given in Section~\ref{sec:overview} and the GIST operator API is
presented in Section~\ref{sec:gist}. Sections~\ref{sec:coref}
and~\ref{sec:lbp} showcase two high-profile machine learning
algorithms and applications that involve large-scale state
transformation and their implementation using GIST and integrated
DBMS system. Finally, we show that the GIST API and a efficient
integration with a DBMS system results in orders-of-magnitude
speedup in Section~\ref{sec:experiment}. }




\section{Query-Driven Entity Resolution Preliminaries}
\label{sec:preliminaries}

In this section we present a foundation of concepts discussed in this chapter.
We start with an introduction of factor graphs then discuss sampling
techniques over this model.
Finally, we formally introduce state-of-the-art entity resolution approaches and explain the origin.

\subsection{Factor Graphs}
\label{sec:factorgraph}
\begin{figure}
  \centering
  \includegraphics[width=0.6\textwidth, trim=3.5cm 3.0cm 0.5cm 4.6cm, clip=true]{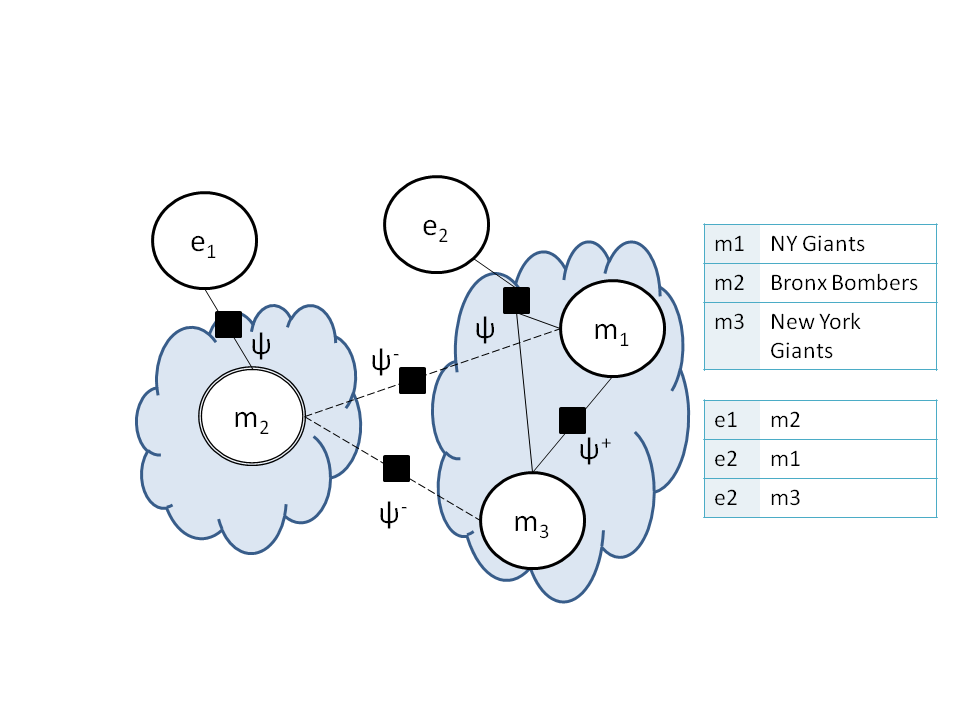}
  \caption{Three node factor graph. Circles (random variables) with $m_i$ represent mentions and those
  with $e_i$ represent entities. Clouds are added for visual emphasis of entity clusters}
  \label{fig:examplefactorgraph}
\end{figure}

{Graphical models} are a formalism for specifying complex probability distributions over many
interdependent random variables.
{Factor graphs} are bipartite graphical models that can capture arbitrary relationships between random
variables through the use of factors~\cite{koller2009probabilistic}.
As depicted in Figure~\ref{fig:examplefactorgraph},
links always connect random variables  (represented as circles) and factor nodes (represented as black squares).
Factors are functions that take as input the current setting of connected
random variables, and output a positive real-valued scalar indicating the
compatibility of the random variables settings.
The probability of a setting to all the random variables is a normalized product of all the factors.
Intuitively, the highest probability settings have variable assignments that yield the highest factor scores.

We use factor graphs to represent complex entity resolution
relationships. Nodes (random variables) may correspond to mentions
of people, places and organizations in documents. Nodes also represent the random variables that
correspond to groups of mentions (entities), these nodes are accompanied by clouds in Figure~\ref{fig:examplefactorgraph}. The factors between
mentions and entities give us a sound representation for
many possible states. The factor graph model also gives us a simple
mathematical expression of the relationship.

Formally, a factor graph $\mathcal{G} = \langle \textbf{x}, \psi \rangle$ contains a set of random
variables $ \textbf{x} = {\{x_i \}}^n_1$ and factors ${\textbf{\boldmath$\psi$\unboldmath}} = { \{\psi_i\} }^m_1 $.
Each factor $ \psi_i$ maps the subset of variables it is associated with to a non-negative compatibility value.
The probability of a setting $\mathbf{\omega}$ among the set of all possible settings $\Omega$ occurring
in the factor graph is given by a probability measure:
\begin{align*}
\pi(\mathbf{\omega}) = \frac{1}{Z} \sum_{x \in \omega} \prod_{i=1}^{m} \psi_i(x^i) &,&  Z = \sum_{\omega \in \Omega} \sum_{x \in \omega}  \prod_{i=1}^{m} \psi_i(x^i) 
\end{align*}
where $x^i$ is the set of random variables that neighbor the factor $\psi_{i}(\cdot)$ and $Z$ is the normalizing constant.

Querying graphical models produces the most likely setting for
the random variables. A query on a factor graph is defined as a
triple $\langle x_q, x_l, x_e \rangle$ where $x_q$ is
the set of nodes in question, $x_l$ is a set of latent nodes (entities) that
are marginalized and $x_e$ is a set of evidence nodes (observed mentions).
A query task is a sum over the all
latent variables and the maximization of the query probability. A
query over the factor graph is defined as
$$ \mathcal{Q}(x_q, x_l, x_e, \pi) = \textrm{argmax}_{x_q} \sum_{v_l \in x_l} \pi(x_q \cup v_l \cup x_e).$$
To obtain the best setting of the queries in question, inference is required.

Several methods exist for performing inference over factor graphs.
The entity resolution factor graph, being pairwise, is dense and highly connected.
This property suggests the best methods for inference are Markov Chain Monte Carlo (MCMC) methods;
in particular, we use a Metropolis Hastings variant~\cite{koller2009probabilistic}.
We refer the reader to our previous work for a detailed discussion on inference over factor graphs and a deviation of the technique~\cite{qam}.


%

\subsection{Inference over Factor Graphs}
\label{sec:inferencefactorgraphs}
Several methods exist for performing inference over
factor graphs.
The entity resolution factor graph, being pairwise, is dense and highly connected.
This property suggests the best methods for inference are Markov Chain Monte Carlo (MCMC) methods;
in particular, we use a Metropolis Hastings variant~\cite{koller2009probabilistic}.

The idea of MCMC-MH is to propose modifications to a current setting and use the
model to decide whether to accept or reject the proposed setting as a replacement for the current settings.
When the models are being scored only the factors touching nodes with changed values, the Markov blanket, needs to be recomputed.
We accept or reject changes so the model can iteratively proceed to an optimal setting.

More formally, consider an MCMC transition function
$ T: \Omega \times \Omega\rightarrow [0,1] $
where given the current setting $\omega$ we can sample a
subsequent setting $ \omega^\prime $.

The probability of accepting a transition given a graphical
model distribution $\pi$ is: 

\begin{equation}
 A(\omega,\omega^\prime) = \textrm{min}\left( 1, \frac{\pi(\omega^\prime)T(\omega,\omega^\prime)}{\pi(\omega)T(\omega^\prime,\omega)} \right). 
\end{equation}

Additionally, the intractable partition function $Z$ is canceled out, making
sample generation inexpensive.
This property allows us to calculate the probability of accepting the next state by simply computing the difference in score between the next and current state~\cite{qam}.

We say the algorithm converges when a steady state is reached.\footnote{We refer to literature for a more detailed description of convergence~\cite{qam}.}
{Intelligently sampling next states decreases the time to convergence.}
Convergence in MCMC is difficult to verify~\cite{cowles1996markov}, we discuss convergence estimation in Section~\ref{sec:experimentsetup}.

\subsection{Cross-Document Entity Resolution}
\label{sec:er}

Cross-document ER is the problem of clustering mentions that appear
across independent sets of documents into groups of mentions that
correspond to the same real world entity. These ER tasks
typically assume a set of preprocessed documents and perform linking
across documents~\cite{bagga1998entity,singh2011large}. The scale of
the cross-document ER problem is typically several
orders of magnitude more than intra-document ER\@. There are no
document boundaries to limit inference scope and all entity mentions
may be distributed arbitrarily across millions of documents.

To model cross-document ER,
let $\mathcal{M} = \{m_1, \ldots, m_{|\mathcal{M}|} \} $ be the set of mentions in a data set.
Each mention $m_i$ contains a set of attribute-value data points.
Let $\mathcal{E} = \{e_1, \ldots, e_{|\mathcal{M}|}\}$ represent the set of entities where each $e_i$
contain zero or more mentions.
Note, we assume the maximum number of entities is no more than the number of mentions and no less than 1.
Each mention may correspond to a unique entity or all mentions may correspond to a single entity.

The baseline method of entity resolution is a straight-forward application of
the MCMC-MH algorithm. We show pseudo code for the baseline method in Algorithm~\ref{algo:erbase}.

\begin{algorithm}
  \textbf{INPUT:} A set of unresolved entities $\mathcal{E}$ each with one mention $m$.\\
  \textbf{INPUT:} A positive integer $samples$.\\
  \textbf{OUTPUT:} A set of resolved entities $\mathcal{E}$.\\
\begin{algorithmic}[1]

\While{{$samples\textrm{-}\textrm{-} > 0$}}
\State $e_i \sim_u \mathcal{E}$ \label{line:erdraw1}
\State $e_j \sim_u \mathcal{E}$ \label{line:erdraw2}
\State $m \sim_u e_i$ \label{line:mentiondraw1}
\State $\mathcal{E^\prime} \leftarrow $\textsc{move}$(\mathcal{E}, m, e_j)$
\If{\textsc{score}($\mathcal{E}$) $<$  \textsc{score}($\mathcal{E^\prime}$)}
\State $\mathcal{E} \leftarrow \mathcal{E^\prime}$
\EndIf
\EndWhile

\Return $\mathcal{E}$
\end{algorithmic}
\caption{The baseline entity resolution algorithm using Metropolis-Hastings sampling}
\label{algo:erbase}
\end{algorithm}

Algorithm~\ref{algo:erbase} takes as input a set of entities $\mathcal{E}$ and
$samples$ which is the number of iterations of the algorithm or a function to estimate convergence.
The algorithm samples two entities from the entity set and moves one random\footnote{Given a set $X$, the function $x\sim_uX$ makes a uniform sample from the set $X$ into a variable $x$.}mention into the other entity.
After the move, the algorithm checks for an improvement in the overall score of the model.
If the model score improves, the changes are kept, otherwise the proposed changes are ignored.
The $\textsc{SCORE}$ function sums the weights of all the edges in
the given entity to obtain a value for the model. This is equivalent
to the probability of the setting $\pi(\cdot)$ as described in
Section~\ref{sec:factorgraph}.

{\bf Blocking.}
Blocking or canopy generation is a preprocessing technique to partition large amounts of data into
smaller chunks, or blocks of items that are likely to be matches~\cite{mccallum2000efficient}.
Blocking can use simple and fast techniques such as sorting based on attributes 
or more advanced techniques that map similar items onto a vector space \cite{DasSarma:2012:ABM:2396761.2398403,5767835, Whang:2009:ERI:1559845.1559870}.


In this chapter, we use two methods of blocking. First, we use an approximate 
string match over all the mentions in the database.
To perform the approximate string filter we use a $q$-grams technique over all the mentions in the database. 
This method creates an inverted index for each mention in the database so a query can be performed to look for all words
that contain a sufficient number of matching $q$-grams. 
This gives us a fast high-recall
filter over many records~\cite{gravano2001using}.

The second is an implicit blocking structure created by computing the influence a
query node has on the other nodes in the data set (see Section~\ref{sec:vose}).
This method uses an estimate of the distant between the query nodes and the candidate mentions to prioritize samples.

\section{Query-Driven Entity Resolution Problem Statement}
\label{sec:problemstatement}

In this section, we formally define the problem of query-driven ER.\@
We use an SQL-like formalism to model traditional and query-driven entity resolution.


In a probabilistic database, let a {Mentions} table
contain all the extracted mentions from a text corpus.
Its column {entity\(\sp{p}\)} represents the $p$robabilistic latent entity labels;
they contain a mapping but that mapping may not represent the current state.
The {People} table holds a watchlist of mentions and relevant contextual information.
The context column is an abstract place holder for text data or richer schemas.
This model only assumes there is a master column, the realization of the
context column is flexible and implementation dependent.

{
\begin{alltt}
Mentions(\underline{docID}, \underline{startpos}, mention, entity\(\sp{p}\), context)
People(\underline{peopleID}, mention, entity\(\sp{p}\), context)
\end{alltt}
}


We also define a user-defined function {coref\_map} that performs
maximum a posteriori (MAP) inference on the
latent entity\(^p\) random variables.
The function takes two instances of mentions with at least one being from a
probabilistic table such as the {Mention} table.
When the query is executed the coref\_map function returns true if the mentions referenced are coreferent.
Following, we describe the traditional exhaustive ER task as well as the
single- and multi-node query-driven ER queries.

\paragraph{Exhaustive} The goals of traditional entity
resolution is to cluster all mentions in a data set. All the
mentions clustered inside each entity are coreferent with each other and not
entity with any mention that is a part of a different entity
cluster. The process of exhaustive ER can be modeled as a
self-join database query where each mention is grouped into coreferent
clusters.
In Algorithm~\ref{alg:exhaustiveq} we create a view displaying the results of a resolved query.

\begin{algorithm}
\begin{lstlisting}[language=SQL,breaklines=true,frame=none,basicstyle=\normalsize, keywordstyle=\ttfamily, identifierstyle=\ttfamily\bfseries, commentstyle=\color{gray},showstringspaces=false]
CREATE VIEW CorefView AS
  SELECT m.docID, m.startpos, m.mention, m2.mention
  FROM Mentions m, Mention m2
  WHERE coref_map(m.*, m.entity^p), m2.mention, m2.context)
\end{lstlisting}
\caption{Example exhaustive entity resolution query that createds a database view}
\label{alg:exhaustiveq}
\end{algorithm}

To obtain unique entity clusters, we can perform an aggregation query over the CorefView.
In Figure~\ref{fig:query_resolved} we see an example of the result of traditional entity resolution.

\paragraph{Single-node Query}
In the ER task, we may only be interested in the mentions
of one entity.
We represent this entity with a template mention, or as a query node $q$.
Single-node entity resolution is modeled as a selection query with a where-clause that
includes the template mention $q$ and returns only the mentions that are members of
the entity cluster that contains the sample mention.
Given a template mention $q$ and its context $q${\tt{.context}}, Algorithm~\ref{alg:singleq} we show the
single-node query based on an example in Section~\ref{sec:selection-driven}.

\begin{algorithm}
\begin{lstlisting}[language=SQL,breaklines=true,frame=none,basicstyle=\normalsize, keywordstyle=\ttfamily, identifierstyle=\ttfamily\bfseries, commentstyle=\color{gray},showstringspaces=false]
SELECT m.docID, m.startpos, m.mention
FROM Mentions m
WHERE coref_map(m.*, m.entity^p), q, q.context)
\end{lstlisting}
\caption{Single query-node driven entity resolution query}
\label{alg:singleq}
\end{algorithm}

Here we add parameters to the coref\_map function that contain the
specific query and its context. It performs ER over the
mentions table but only returns an affirmative value if the labels
for the entity cluster match the query node. For example, if the
template mention $q$ was `Mark Zuckerberg', and the query context
were keywords such as `facebook' and `ceo', the only returned
mentions will be those that represent {Mark Zuckerberg} the
facebook founder. This is similar to a `facebook' approximate string
search.
The emphasis of this chapter is optimizing this function so while performing ER we perform 
less work compared an exhaustive query.


\paragraph{Multi-Query}
In many cases, a user may be interested in a watchlist of entities.
Watchlist is a subset of the larger Mention set.
This is common for companies looking for mentions of its products in a data set.
In this case, mention are only clustered with the entities represented in the watchlist.
Algorithm~\ref{alg:joinq} is an example of a join-query between the {Mentions} table and the {People} table.

\begin{algorithm}
\begin{lstlisting}[language=SQL,breaklines=true,frame=none,basicstyle=\normalsize, keywordstyle=\ttfamily, identifierstyle=\ttfamily\bfseries, commentstyle=\color{gray},showstringspaces=false]
SELECT m.docID, m.startpos, m.mention, q
FROM Mentions m, People q
WHERE coref_map(m.*, m.entity^p), q, q.context)
\end{lstlisting}
\caption{Multi-query between the Peoples watch list tablse and the full mentions set}
\label{alg:joinq}
\end{algorithm}

This function combines a watch list of terms and performs ER with respect to
the specific examples in the watchlist.
The multi-query method uses scheduling to perform inference, or a fuzzy equal, over each mentions.
In Section~\ref{sec:jointquery} we propose scheduling algorithms so multi-query node 
ER gracefully manage multiquery workloads.


\eat{%
\begin{figure*}
    \centering
    \begin{subfigure}[b]{0.25\textwidth}
        \centering
        \includegraphics[width=.6\textwidth,trim=15.5cm 0cm 0cm 0cm, clip=true,height=.17\textheight]{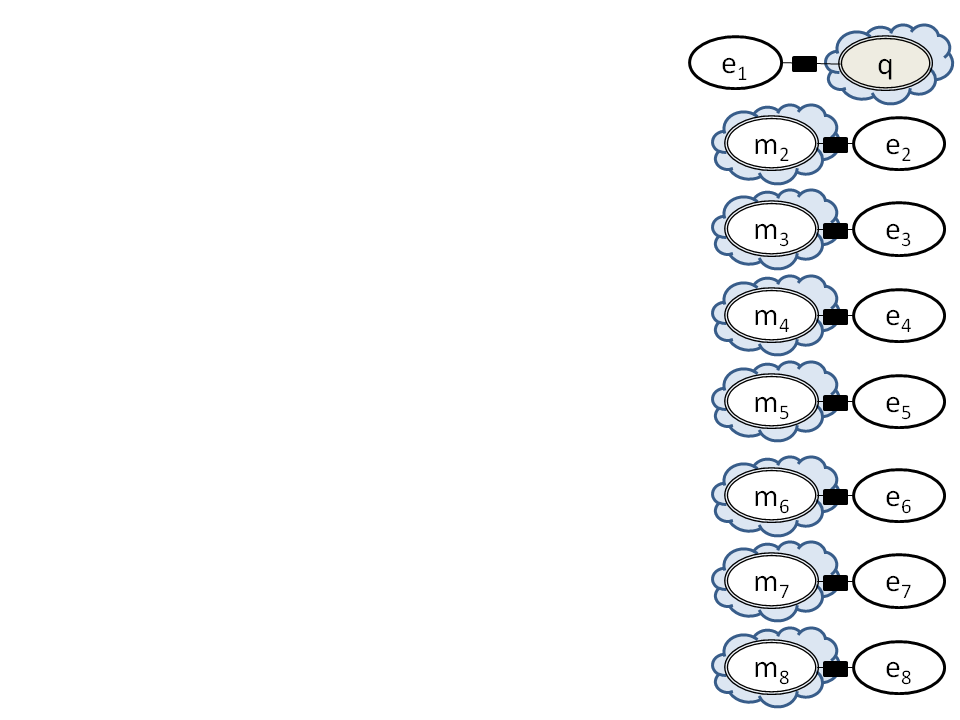}
        \caption{A possible initialization for entity resolution}
        \label{fig:query_start}
    \end{subfigure}
~
    \begin{subfigure}[b]{0.25\textwidth}
        \centering
        \includegraphics[width=.6\textwidth,trim=15.5cm 0cm 0cm 0cm, clip=true,height=.17\textheight]{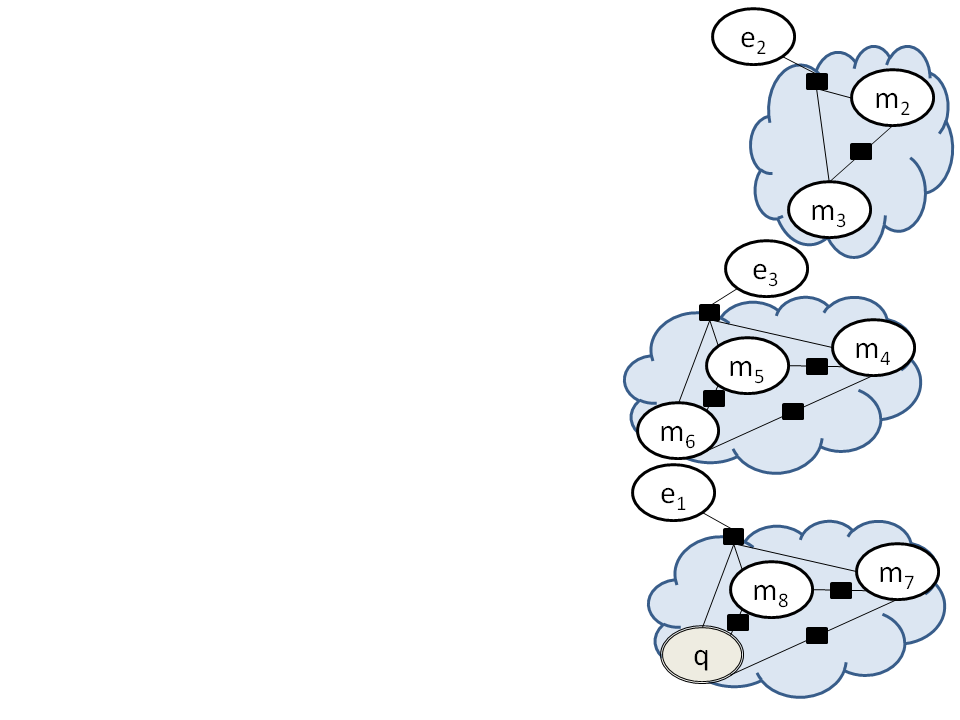}
        \caption{The correct entity resolution for all mentions}
        \label{fig:query_resolved}
    \end{subfigure}
~
    \begin{subfigure}[b]{0.25\textwidth}
        \centering
        \includegraphics[width=.6\textwidth,trim=15.5cm 0cm 0cm 0cm, clip=true,height=.17\textheight]{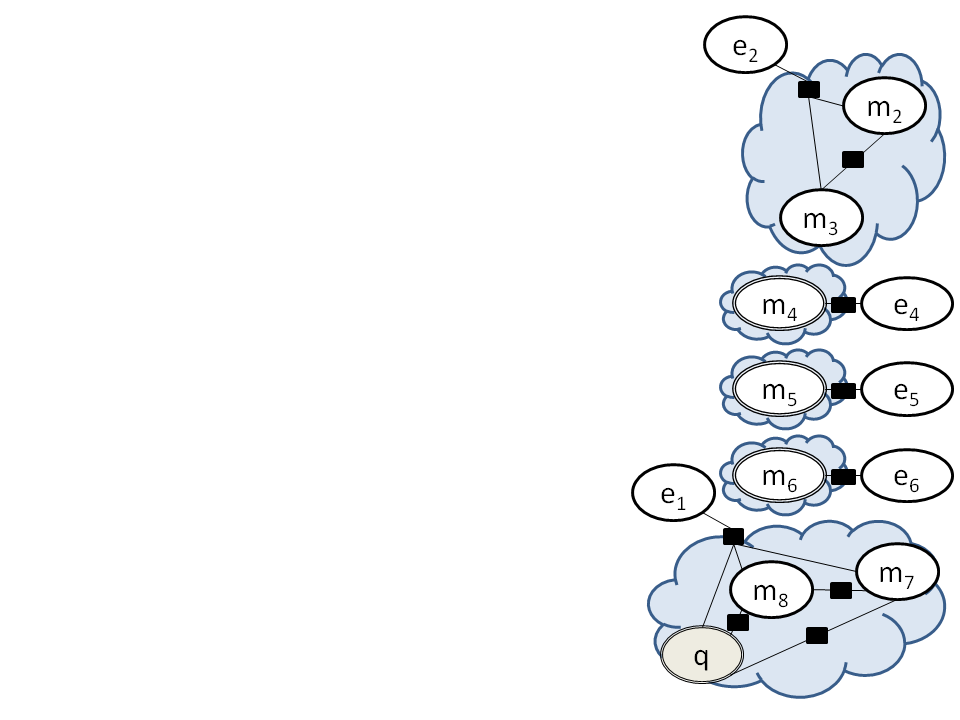}
        \caption{The entity containing $q$ is internally coreferent; the other entities are not correctly resolved}
        \label{fig:query_driven_resolved}
    \end{subfigure}

    \caption{Initialization and possible final arrangements for query-driven entity resolution.}
    \label{fig:factorgraphser}
\end{figure*}
}

\begin{figure}
    \centering
    \includegraphics[width=0.16\textwidth,trim=15.5cm 0cm 0cm 0cm, clip=true,height=.22\textheight]{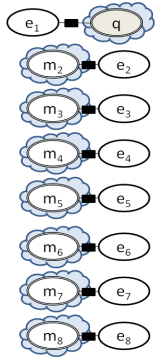}
    \caption{A possible initialization for entity resolution}
    \label{fig:query_start}
\end{figure}
\begin{figure}
    \centering
    \includegraphics[width=0.16\textwidth,trim=15.5cm 0cm 0cm 0cm, clip=true,height=.22\textheight]{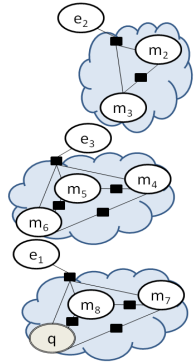}
    \caption{The correct entity resolution for all mentions}
    \label{fig:query_resolved}
\end{figure}
\begin{figure}
   \centering
    \includegraphics[width=0.16\textwidth,trim=15.5cm 0cm 0cm 0cm, clip=true,height=.22\textheight]{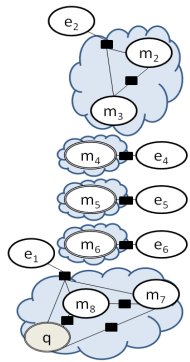}
    \caption{The entity containing $q$ is internally coreferent; the other entities are not correctly resolved}
    \label{fig:query_driven_resolved}
\end{figure}


\section{Query-Driven Entity Resolution Algorithms}
\label{sec:algorithms}


Query-driven ER is an understudied problem; in this section
we describe our approach to query-driven ER with one entity (single-query ER)
and with multiple entities (multi-query ER).
First, we give a graphical intuition of query-driven ER algorithms.

\subsection{Intuition of Query-Driven ER}

In this section, we remind the reader of the query-driven ER task with a formal definition.\@
Each ER task is given a corpus $\mathcal{G}$ and a set
of entity mentions $\mathcal{M} = \{m_1, \ldots, m_{|m|}\}$ extracted from the $\mathcal{G}$.
A user may supply set of query nodes $Q =\{q_1, \ldots, q_{|Q|}\}$.
Each $q_i$, also called a query template, may be a member of $\mathcal{M}$ or a 
manually declared mention that is appended to the set of mentions.
For each node $q_i \in Q$, the task of ER is to compute the set of mentions $E = \{e_1, \ldots, e_{|Q|}\}$ that 
only contain mentions that are coreferent with the query node,
\[e_{q_i} = \{m_i | m_i \in \mathcal{M}, \texttt{QDER}(\mathcal{M}, m_i, q_i)\}.\]
In Section~\ref{sec:selection-driven}, we describe implementations of the \texttt{QDER} algorithm for $|Q| = 1$.
In Section~\ref{sec:jointquery}, we describe techniques of scheduling the ER task for the general case of $|Q| > 1$.

Fundamentally, the ER algorithm generates a graphical model and makes new state proposals (jumps) to reach the best state (see Section~\ref{sec:preliminaries}).
The query-driven algorithms in this section use a query node to facilitate more sophisticated jumps.
By making smart proposals we expect faster convergence to an accurate state.
As a note to the reader, a summary of query-driven algorithms can be found in Table~\ref{tab:algorithmsummarytable}.

Figure~\ref{fig:query_driven_resolved} shows an initial configuration and acceptable query-driven entity resolution solutions.
An example initial state of this algorithm is shown in Figure~\ref{fig:query_start} --- each mention is
initially assigned to separate entities.
Alternatively, the model may be initialized randomly, or in an arrangement from a
previous entity resolution output or with all mentions in one entity.
Figure~\ref{fig:query_resolved} is the full resolution for the data set;
each mention is correctly assigned to its entity cluster.
Figure~\ref{fig:query_driven_resolved} is a result that was resolved with query-driven methods and is a partially resolved data.
Because the entity containing the query node is completely resolved the solution is acceptable.

\subsection{Single-Node ER}
\label{sec:selection-driven}
Single-node ER algorithms are the class of algorithms that resolve a single query-node as discussed in Section~\ref{sec:problemstatement}.
In particular, the
{target-fixed} ER algorithm aims to focus a majority of the proposals on resolving the query entity.
The algorithm fixes the query node as the target entity and then randomly
selecting a source node to merge into the entity of the target query node.
This focus on building the query entity in this type of importance sampling means
the query entity should be resolved faster than if we sampling each entity uniformly.

A query-driven ER algorithm that only selects the query-node as the target entity during sampling will create errors
because such an algorithm is unable to remove erroneous mentions from the query entity.
To prevent these errors, we allow the algorithm to occasionally back out of poor decisions, that is, it makes
non-query specific samples.
Shown in Algorithm~\ref{algo:qderfixed}, {target-fixed} entity resolution adapts Algorithm~\ref{algo:erbase}
but it allows parameters to specify the proportion of time the different sampling methods are selected.

In addition to the input mentions $\mathcal{E}$
from Algorithm~\ref{algo:erbase}, {target-fixed} entity resolution takes as input a
query node $q$.
The output of the algorithm is a resolved query entity and other
partially resolved entities.

For each sampling iteration the algorithm can make two decisions.
The sampler may propose to merge a random
source node that is not already a member of the query entity into
the target query entity.
Alternatively, the algorithm merges a random node with a random entity.

\begin{algorithm}
  \alginput{A query node $q$.\\
  A set of entities $\mathcal{E}$ each with one mention $m$.\\
  A positive integer $samples$.}\\
  \algoutput{A set of resolved entities $\mathcal{E^\prime}$.i}\\
\begin{algorithmic}[1]

\State $\mathcal{E^\prime} \leftarrow \mathcal{E} \cup q$
\While{$samples\textrm{-}\textrm{-} > 0$}

\If{$\textrm{\textsc{random}}() < \tau_\alpha$} \label{algo:qderfixed:targetfixedstart}

    \State $e_i \sim_u \mathcal{E^\prime}$
    \State $e_j \leftarrow q.entity$
    \State $m \sim_u e_i$ \label{algo:qderfixed:targetfixedend}

\Else \label{algo:qderfixed:randomstart}

    \State $e_j\leftarrow \{e | \exists e, e \in \mathcal{E^\prime}, e \not= q\textrm{.entity}\}$
    \State $e_i\leftarrow \{e | \exists e, e \in \mathcal{E^\prime}, e \not= e_j\}$

    \State $m \sim_u e_i$ \label{algo:qderfixed:randomend}
\EndIf

\State $\mathcal{E^{\prime\prime}} \leftarrow $\textsc{move}$(\mathcal{E^\prime}, m, e_j)$ \label{algo:qderfixed:mergestart}
\If{\textsc{score}($\mathcal{E^\prime}$) $<$ \textsc{score}($\mathcal{E^{\prime\prime}}$)}
    \State $\mathcal{E^\prime} \leftarrow \mathcal{E^{\prime\prime}}$
\EndIf \label{algo:qderfixed:mergeend}

\EndWhile

\Return $\mathcal{E^\prime}$
\end{algorithmic}
\caption{\footnotesize Target-fixed entity resolution algorithm}
\label{algo:qderfixed}
\end{algorithm}

On lines~\ref{algo:qderfixed:targetfixedstart} to~\ref{algo:qderfixed:targetfixedend} the algorithm
takes a uniform sample from the list of entities. If the sampled entity is the same as the query
entity it tries again and samples a distinct entity. A node is drawn from this entity.
The probability of this block being entered is $\tau_\alpha$.
Lines~\ref{algo:qderfixed:randomstart} to~\ref{algo:qderfixed:randomend} are entered with a
probability $(1 - \tau_\alpha)$.
This block performs a random entity assignment in the same manner as Algorithm~\ref{algo:erbase}.
This block offsets the aggressive nature of the target-fixed algorithm by probabilistically backing out of
any bad merges.
Finally, the block starting from line~\ref{algo:qderfixed:mergestart} to line~\ref{algo:qderfixed:mergeend}
scores the new arrangement and accepts if this improves the model score.
We discuss parameter settings in Section~\ref{sec:summarydiscussion}.

\begin{table}
\centering
\caption{Mentions sets $\mathcal{M}$ from a corpus}
\begin{tabularx}{\textwidth}{XXXX}
\hline
{id} & {Mention} & {\ldots} \\
\hline
$m_1$ & NY Giants & \ldots \\
$m_2$ & Bronx Bombers& \ldots \\
$m_3$ & New York Giants & \ldots \\
$m_4$ & Yankees &  \ldots \\
$m_5$ & Brooklyn Dodgers & \ldots \\
$m_6$ & The Yanks & \ldots \\
\hline
\end{tabularx}
\label{tab:examplemention}
\end{table}

\begin{table}
\centering
\caption{Example query node $q$}
\begin{tabularx}{\textwidth}{XXX}
\hline
{id} & {Mention} & {\ldots}\\
\hline
$q$ & New York Yankees & \ldots \\
\hline
\end{tabularx}
\label{tab:examplequery}
\end{table}

\paragraph{Example}
Take the synthetic mention set $\mathcal{M}$ shown in
Table~\ref{tab:examplemention} and a query node $q$, the baseball
team `New York Yankees', in
Table~\ref{tab:examplequery}.
This is the result of the approximate match of query $q$ over a larger data set (blocking).
The mentions of $\mathcal{M}$ may be initialized by assigning each
mention to its own entity.
After a successful run of traditional entity resolution the set
of entities clusters are
$$\{\langle q, m_2, m_4, m_6\rangle, \langle m_1, m_3 \rangle, \langle  m_5 \rangle \}.$$
For query-driven scenario
the only entity we are interested in is $\langle q, m_2, m_4, m_6\rangle$.
Each mention in this query entity is an alias for the `New York Yankees' baseball team.
The other two mentions represent the `New York Giants' football team and
the `Brooklyn Dodgers' baseball team respectively.

The {target-fixed} algorithm attempts to merge nodes with the query entity
one mention at a time and the merge is accepted if it improves the score
of the overall model.
We can see in the example that a merge of $m_1$ and $m_3$
may improve the overall model because they have similar keywords
but one refers to the query entity and the other to different football team.
The {target-fixed} algorithm can correct this type of error by probabilistically backing out
of errors by moving mentions in the query node to a
new entity as show in line~\ref{algo:qderfixed:randomstart} to line~\ref{algo:qderfixed:randomend} of Algorithm~\ref{algo:qderfixed}.


\subsection{Multi-query ER}
\label{sec:jointquery}

A user may want to resolve more than one query entity, that is, she may be interested in resolving a watch list of entities
over the data set.
To support multiple queries, first
merge the canopies of each query node in the watch list to obtain a subset of the full graphical
model containing only the nodes similar to query nodes.
To resolve the entities we can use query-proportional methods iteratively over each query node.
We define two classes of schedules, namely, static and dynamic.

Static schedules are formulated before sampling while dynamic
schedules are updated in response to estimated convergence. The two
static schedules we develop are {random} and
{selectivity-based}. In random scheduling each query
node from the watch list is selected in a round robin style.
Selectivity-based scheduling is a method of ordering multi-query samples to schedule proposals in proportion to the
selectivity of the query node.
Selectivity, in this case, is defined as the number of mentions retrieved using an approximate
match of the data set, or the query node's contribution to the total new graphical model.
For example, the selectivity of our query node $q$ in Table~\ref{tab:examplequery} the selectivity
is simply the size of $\mathcal{M}$, shown in Table~\ref{tab:examplemention}.

Random-based scheduling method performs well if all query nodes come from similar selectivity.
Otherwise, if the selectivity of each query node vary,
one query node may require more sampling compared to the others.
If one query node needs a lot of samples to converge,
it may take the whole process a long time to complete and cycles may
be wasted on other nodes that have already converged.

In addition to scheduling samples in proportion to their
selectivity, we can schedule samples dynamically, depending on the
progress of each query entity.
To perform dynamic scheduling we need to know how each query entity is progressing towards convergence.
To estimate the running convergence we do not use standard techniques in literature
because scheduling needs to occur before the model is close to convergence~\cite{cowles1996markov}.
Instead, we estimate the convergence by measuring the fraction of accepted samples over the
last $N$ samples of each query in the watch list.
The two dynamic scheduling algorithms are {closest-first} and sampling the {farthest-first}.
In {closest-first} we queue up the query node that has the lowest positive
average number of accepted nodes over the last $N$ proposals.
This scheduling method performs inference for the node that is
closest to being resolved so it can move on to other nodes.
Alternatively, the {farthest-first} algorithm schedules the node that has the highest convergence rate.
This scheduling algorithm makes each query entity progress evenly.


\section{Optimization of Query-Driven ER}
\label{sec:queryproportional}

The previous ER techniques aggressively attempt to resolve the query entity.
However, if the query node is not representative of the query items performance of target-fixed ER can lead to undesirable results.
We do not explore this trade-off; we assume users can select representative query nodes.
In this section, we introduce optimizations to create approximate query-driven
samples based on the query node.
We first discuss the influence function that is used to make query-driven
proposals. 
We then discuss the {attract} and {repel} versions of the influence function followed by
two new algorithms.
We end with implementation details and a summary of our query-driven algorithms.

\subsection{Influence Function: Attract and Repel}
\label{sec:vose}

To retrieve nodes from a graphical model that is similar to a query node we employ the notion of influence.
Our assumption is that nodes that are similar have a high probability of being coreferent.
An {influence trail score} between two nodes in a graphical model
can be computed as the product of factors along their active trail as defined
in literature~\cite{qam}.
For a node $m_i \in \mathcal{M}$ and the query node $q \in \mathcal{M}$ the influence of $m_i$ on the query node
is defined as:
$$ \mathcal{I}(m_i, q) = \sum_{j \in \mathcal{F}} w_j \psi_j(m_i,q)$$
where $\mathcal{F}$ is the world of pairwise features and
the feature weight and log-linear function
are, respectively,
$w_j$ and $\psi_j$.
The influence function $\mathcal{I}$ is an implementation of this trail score.

The influence function takes a set of entities --- or the equivalent GM --- and a query node $q$ as parameters.
The parameters to an influence function can be over the whole database or a canopy.
Over several invocations of the function, $\mathcal{I}$ returns mentions
from the graphical model with a frequency proportionate to their influence on $q$.
If a mention has little or no influence, the influence acts as a blocking function, infrequently returning the mention.
Recall influence is the distance active trail distance to query node.
To implement the influence function we build a data structure based on an
algorithm by Vose~\cite{Vose:1991:LAG:126262.126280}, hereafter referred to as a Vose structure.

The input mentions to the blocking algorithms may result in high or low quality canopies.
A high quality canopy means
most of the mentions in the canopy are associated with the query node.
Low quality canopies, which are more common, corresponds to only a small number of mentions
being associated with the query node.
When initializing query-driven algorithms the canopy quality is important
for determining what algorithm to use.

The {attract} method initializes each mention in the canopy in its own entity, and then
mentions are merged until the convergence.
The target-fixed algorithm discussed in Section~\ref{sec:selection-driven} is explained using this method.
The {attract} method works well for low quality canopies, or canopies
that require a small number or items to merge.
Conversely, the {repel} method works well with high quality canopies or
when most items in a canopy belong to the query entity.

The {repel} method initializes each mention in the canopy into a single entity.
Then proposals are made to remove mentions from the entity so we are left with
only the nodes in the query entity.
We discuss this method using the hybrid algorithm in Section~\ref{sec:coinfluence}.
To build an influence function for the repel method
we can use the same method and we only need to normalize and invert the influence scores.
We refer to this as co-influence or $\bar{\mathcal{I}}$.

\subsection{Query-proportional ER}
\label{sec:queryproper}

In the {query-proportional} sampling algorithm, on every iteration,
the source mention and target entity are selected in proportion
to its distance to the query entity.
Instead of focusing solely on the query entity, this algorithm prioritizes
samples using a measure that represents probability of a mention being coreferent with the query entity.

That is, each node $p$ in the graphical model $\mathcal{G}$ is selected on the {active trail} between
itself and the query node $q$.
This algorithm merges nodes that are similar to the query node with an increased frequency.

Before query-proportional sampling, a data structure for $\mathcal{I}$ is created.
The $\mathcal{I}$ influence structure takes a query node $q$ and the global graphical model $\mathcal{E}$ then
returns a sampled mention. As $\mathcal{I}$ is called multiple times, the distribution of the nodes returned
is proportional to their influence.
Algorithm~\ref{algo:queryproportional} describes the query-proportional algorithm.

\begin{algorithm}
  \alginput{A query node $q$ to drive computation.\\
  A set of entities $\mathcal{E}$ each with one mention $m$.\\
  A positive integer $samples$.\\
  A function $\mathcal{I}$ that samples from nodes entities according to its influence on a mention.}\\
  \algoutput{A set of resolved entities $\mathcal{E^\prime}$.}\\
\begin{algorithmic}[1]

\State $\mathcal{E^\prime} \leftarrow \mathcal{E} \cup q$
\While{$samples\textrm{-}\textrm{-} > 0$}

\State $m_1 \leftarrow \mathcal{I}(\mathcal{E^\prime}, q)$ \label{algo:queryproportional:samp1}
\State $m_2 \leftarrow \mathcal{I}(\mathcal{E^\prime}, q)$ \label{algo:queryproportional:samp2}

\State $\mathcal{E^{\prime\prime}} \leftarrow $\textsc{move}$(\mathcal{E^\prime}, m_1, m_2.entity)$
\If{\textsc{score}($\mathcal{E^\prime}$) $<$ \textsc{score}($\mathcal{E^{\prime\prime}}$)}
\State $\mathcal{E^\prime} \leftarrow \mathcal{E^{\prime\prime}}$
\EndIf
\EndWhile

\Return $\mathcal{E^\prime}$
\end{algorithmic}
\caption{\footnotesize Query-proportional algorithm}
\label{algo:queryproportional}
\end{algorithm}

For each iteration, the algorithm selects mentions using the
influence function (line~\ref{algo:queryproportional:samp1} and line~\ref{algo:queryproportional:samp2}).
Then, one mention $m_1$ is moved into the entity of $m_2$.
Mentions $m_1$ and $m_2$ have a higher probability of being coreferent and therefore
a higher probability of a merge occurring in the query entity compared to random selections as in Algorithm~\ref{algo:erbase}.
As a corollary, the influence sampling property creates many small entities
that are similar to the query entity.

\eat{We can see a with an explanation that mentions that are similar to
  the query node are proposed for a merge more frequently.
  Let sequences $A = \{a_1, \ldots, a_n \}$ and $B = \{b_1, \ldots, b_n \}$ be
  ordered sequences that map to entities in a data set $\mathcal{G}$.
  Let the value of $a_i$ and $b_i$ be proportional to the distance from
  some query node where $a_i \geq a_{i+1}$ and $b_i == b_{i+1}$.
  That is, the values in each sequence correspond to the probability the value will be sampled from the sequence.
  For example, if $a_2 = \frac{1}{4}$ the probability item $a_2$ will be selected is $0.25$.
  Let $A$ be the distribution associated with the query-proportional
  influence function and all values in $B$ have a uniform value.

  So for some positive constant $L$ the sum of each sequence can be described as
  $ \sum_i^n a_i = \sum_i^n \frac{1}{L (n-i)^2 } = 1$ and
  $ \sum_i^n b_i = \sum_i^n \frac{1}{n} = 1$ and for a sufficiently large $n$
  the two sums are equal.
  The expected value of a proportional sample over the values in $A$ will be an item at a lower index
  than a sample from sequence $B$. In other words, the distribution mean of $A$ will have a
  positive skew compared to the $B$.
  The positive skew explains the mean of items sampled are more similar to the query node compared to
  the uniform mean. Therefore, the average samples from $A$ are closer in influence to the query node.
  Closer to the influence means nodes are similar and the probability of a merge occurring is greater.
  }

During query-proportional sampling more entities that are similar to the query node are created.
Some of the mentions created in intermediate entities during query-proportional sampling will move to the
query entity.
This is a big advantage when performing entity-to-entity merges (as opposed to
mention to entity merges). In this chapter, we do not investigate this extension to
the algorithm.



\subsection{Hybrid ER}
\label{sec:coinfluence}

The best of both the target-fixed and query-proportional algorithms can be
combined to create a {hybrid algorithm}.
Like the target-fixed algorithm, the hybrid method aggressively fixes the target as
the query entity.
The hybrid method also chooses its source node using the influence function in the same manner as the
query-proportional algorithm.





Algorithm~\ref{algo:cohybrid} shows the hybrid algorithm using the repel method.
With probability $\tau_\alpha$ the algorithm chooses a mention using the repel method
($\bar{\mathcal{I}}$) and moves it to
an entity that is not the query node.
This is the opposite of merging a node into the query entity.
Pseudocode is listed on lines~\ref{algo:cohybrid:samp1} to line~\ref{algo:cohybrid:samp2}.

\begin{algorithm}
  \alginput{A set of entities $\mathcal{E}$, where one contains all the mentions $m$ and the others are empty.\\
  A positive integer $samples$.\\
  A query node $q$.\\
  A function $\overline{\mathcal{I}}$ that samples from nodes entities according to its influence on a mention.}\\
  \algoutput{A set of resolved entities $\mathcal{E^\prime}$.}\\
\begin{algorithmic}[1]

\State $\mathcal{E^\prime} \leftarrow \mathcal{E} \cup q$
\While{$samples\textrm{-}\textrm{-} > 0$}

  \If{$\textrm{\textsc{random}}() < \tau_\alpha$}\label{algo:cohybrid:samp1}

    \State $m \leftarrow \overline{\mathcal{I}}(\mathcal{E^\prime}, q)$
    \State $e_i\leftarrow \{e | \exists e, e \in \mathcal{E^\prime}, e \not= q\textrm{.entity}\}$\label{algo:cohybrid:samp2}

  \Else

    \State $e_i\sim_u \mathcal{E^\prime}$
    \State $e_j\leftarrow \{e | \exists e, e \in \mathcal{E^\prime}, e \not= e_i\}$
    \State $m \sim_u e_j$

  \EndIf
    \State $\mathcal{E^{\prime\prime}} \leftarrow $\textsc{move}$(\mathcal{E^\prime}, m, e_i)$
  \If{\textsc{score} ($\mathcal{E^\prime}$) $ < $ \textsc{score}($\mathcal{E^{\prime\prime}}$)}
    \State $\mathcal{E^\prime} \leftarrow \mathcal{E^{\prime\prime}}$
  \EndIf

\EndWhile

\Return $\mathcal{E^\prime}$
\end{algorithmic}
\caption{\footnotesize Hybrid-Repel algorithm}
\label{algo:cohybrid}
\end{algorithm}



\subsection{Implementation Details}
\label{sec:parallelqder}
The previous algorithms described single process sampling over the set of mentions.
The multi-query methods are modeled for several interwoven sequential single-node ER processes.
In this section, we describe our implementation of the hybrid algorithm over a
parallel database management system.

An independent Vose structure ($\mathcal{I}$,\S~\ref{sec:vose}) is created for each query node in the query set.
The creation of the Vose structure query nodes is parallelized.
When the number of query nodes increases the Vose structures demand more memory from the system.
Each Vose structure contains array of type double precision and unsigned int.
The space for the structure is $O(|Q|\cdot|M|)$ where $|Q|$ is the number of query nodes in the
query and $|M|$ is the number of mentions in the corpus.
The Vose structure is accessed over every sample and needs to be in memory.
To increase scalability, one could store the full sets of precomputed samples and serialize the
Vose structures to disk but that is not explored here~\cite{Jampani:2008:MMC:1376616.1376686}.

Sampling over the query nodes for each algorithm can also be perform in parallel.
In our method, a thread selects a query node using a random schedule as described in Section~\ref{sec:jointquery}.
The system will use the Vose structure associated with the query node to set up a proposal move.
The system attempts to obtain a locks for both entities involved in the proposal.
If the system is unable to obtain a lock on either of the two entities the system will back out
and resample new entities.
When the number of query nodes is small the query-driven algorithms experience lot of contention
at the entities containing the query nodes.
In these circumstances, the system will back out and either restart the proposal process or attempt a baseline proposal.
This avoids waiting for locked entities and keeps the sampling process active.
In Section~\ref{sec:parallelexperiments} we demonstrate the parallel hybrid method over a large data set.

\subsection{Algorithms Summary Discussion}
\label{sec:summarydiscussion}

Algorithms~\ref{algo:qderfixed},~\ref{algo:queryproportional} and~\ref{algo:cohybrid}
are modifications of proposal jumps found in the baseline Algorithm~\ref{algo:erbase}.
Table~\ref{tab:algorithmsummarytable} describes the proposal process for each algorithm by its
preferred jump method.

\begin{table}[h]
\begin{center}
\caption{Summary of algorithms and their most common methods for
proposal jumps}
\begin{tabularx}{\textwidth}{XXX}
\hline
& {source} & {target}\\
\hline
Baseline & random & random\\
Target-Fixed & random & fixed\\
Query-Proportional & proportional & proportional\\
Hybrid & proportional & fixed\\
\hline
\end{tabularx}
\label{tab:algorithmsummarytable}
\end{center}
\end{table}

The target-fixed algorithm builds the query entity by aggressively proposing random samples
to merge into the query entity.
The query-proportional algorithm uses an influence function to ensure its samples are mostly
related to the query node.
The hybrid algorithm mixes the aggressiveness of the target-fixed with the intelligent
selecting of the source node found in the query proportional method.

After choosing the correct algorithm, a user needs to have a well trained model with several features.
An advantage of using query-proportional techniques, because so little sampling is required,
is that we can interactively test query accuracy.
We can and also add context or keywords that were discovered from a previous run of the algorithm.
This interactive querying workflow will help improve accuracy, which we experimentally verify in Section~\ref{sec:experiments}.

\paragraph{Parameter settings}
The algorithm takes several parameters that affect performance.
While not studied in this chapter, parameter settings are robust to change making parameter selection simple.
The first is the number of proposals ($samples$).
This number can be a function on the size of the data set.
Each query node should have the opportunity to be merged into an entity more than once.

The value $\tau_\alpha$ is between $[0.0, 1.0]$ and represents how often to perform the main type of sampling.
This value should be set to a high value, 0.9 for {accept} algorithms.
With probability $1-\tau_\alpha$ the algorithms
back-off to random samples to improve mixing.
This value is lowered to counter some of the aggression, particularly in Algorithm~\ref{algo:qderfixed}.
The parallel experiments use a $\tau_\alpha = 1$ and back out when there is contention in the threads.

In statistics, a negative binomial function is used to model the number of trials it takes for an event to be a success.
We can also use a negative binomial function as a decay function for the output of the influence function.
We use this function because we want values that are most similar (lowest score) to
be sampled more often.
We set the $r$ value, or number of failures for the negative binomial function to 1.
We set the $p$ value, or the probability of each success to a value close to 0.05.

In the multi-query ER algorithms we run inference for $K$ steps before we look to change the query entity.
In our experiments
we choose a $K$ of 500 and an increasing value from two to 100 thousand in the parallel experiments.

\section{Query-Driven Entity Resolution Experiments}
\label{sec:experiments}
In this section, we describe the implementation details, the data sets and our experimental setup.
Next, we discuss our hypotheses and four corresponding experiments. 
We then finish with a discussion of the results.

\paragraph{Implementation}
We developed the algorithms described in Section~\ref{sec:algorithms} in Scala 2.9.1 using the Factorie package.
Factorie is a toolkit for building imperatively defined factor graphs~\cite{mccallum09:factorie}.
This framework allows a templated definition of the factor graoh to avoid fully materializing the structure.
The training algorithms are also developed using Factorie.
The algorithms for canopy building and approximate string matching are developed as inside of PostgreSQL 9.1 and
Greenplum 4.1 using SQL, PL/pgSQL and PL/Python.
Inference is performed in-memory on an Intel Core i7 processors with 3.2GHz, 8 cores and 12GB of RAM\@.
The approximate string matching on Greenplum is performed on a AMD Opteron 6272 32-core machine with 64 GB\@.

The parallel experiments were developed entirely in a parallel database, DataPath~\cite{Arumugam:2010:DSD:1807167.1807224}.
DataPath is installed on a 48-core machine with 256 GBs.

\paragraph{Data sets.}
The experiments use three data sets,
the first is the English newswire articles from the Gigaword Corpus,
we refer to this as the NYT Corpus~\cite{graff2007ldc2007t07}.
The second is a smaller but {fully-labeled} Rexa data set.\footnote{\url{http://cs.neiu.edu/\~{}culotta/data/rexa.html}}
Because it is fully-labeled it allows us to run the more detailed micro benchmarks.
The NYT corpus contains 1,655,279 articles and 29,866,129 paragraphs from the years 1994 to 2006.
We extracted a total of 71,433,375 mentions using the natural language toolkit
named entity extraction parser~\cite{nltk09}.
Additionally, we compute general statistics about the corpus including
the term and document frequency and tf-idf scores for all terms.
We manually labeled mentions for each query over the NYT data set.

The second data set, Rexa, is citation data from a publication search engine named Rexa.
This data set contains 2454 citations and 9399 authors of which 1972 are labeled.
We perform experiments on the Rexa corpus because it is fully labeled unlike the NYT Corpus.
The Rexa corpus is smaller in total size but it has average sized canopies.

The third data set is the Wikilinks Corpus~\cite{singh12:wiki-links} largest labeled corpus for entity resolution that we could find at the time of development.
It contains 40 million mentions and 3 million entities that were extracted from the web and truthed based on web anchor links to Wikipedia pages.
We loaded a million mentions onto DataPath to demonstrate the parallel capabilities.

\subsection{Experiment Setup}
\label{sec:experimentsetup}

Table~\ref{table:features} lists the features and the weights for each feature.
\paragraph{Features.}
Features that look for similarity between mention nodes are called affinity features and they
are given positive weights.
Features that look for dissimilarity between mentions nodes are called repulsion
factors and they are given negative weights.
We implement three classes of features: pairwise token features, pairwise
context features and entity-wide features.
Pairwise features directly compare tokens strings on attributes such as equality or
matching substrings.
Context features compare the information surrounding the mention. We can look at the surrounding sentence,
paragraph, document or user specified keywords.
The query nodes are extracted from text and contain a proper document context.
With this context, we use a tf-idf weighted cosine similarity score
to compare the context of each mention token.
Finally, entity-wide features use all mentions inside an entity cluster to make a decision.
An example entity-wide feature counts the matching mention strings between two entities.


\begin{table}
\begin{center}
\footnotesize
\caption{
Features used on the NYT Corpus.
The first set of features are token specific features,
the middle set are between pairs of mentions
and the bottom set are entity wide features.}
\begin{tabularx}{\textwidth}{XXXX}
\hline
{Feature name} & {Score$^+$} & {Score$^-$} & {Feature type}\\
\hline
Equal mention Strings & +20&-15  & Token Specific\\
Equal first character & +5&  & Token Specific\\
Equal second character & +3&  & Token Specific\\
Equal second character & 0 \\

Unequal mention Strings & & -15 & Token Specific\\
Unequal first character & 0 \\
Unequal second character & 0 \\
Unequal second character & 0 \\

Equal substrings & +30&-150  & Token Specific\\
Unequal substrings & & -150 & Token Specific\\

Equal string lengths & +10&  & Token Specific\\

      Matching first term & +90& -3 & Token Specific\\
   No matching first term & & -3& Token Specific\\

Similarity $\geq 0.99$ & +120& & Pairwise\\
Similarity $\geq 0.90$ & +105& & Pairwise\\
Similarity $\geq 0.80$ & +80& & Pairwise\\
Similarity $\geq 0.70$ & +55& & Pairwise\\
Similarity $\geq 0.60$ & +35& & Pairwise\\
Similarity $\geq 0.50$ & +15& & Pairwise\\
Similarity $\geq 0.40$ & & -5& Pairwise\\
Similarity $\geq 0.30$ & & -50& Pairwise\\
Similarity $\geq 0.20$ & & -80& Pairwise\\
Similarity $< 0.20$ & & -100& Pairwise\\

Matching terms & +20&  & Pairwise\\
         Token in context & +1& & Pairwise\\
      No matching keyword &+700 & -10& Pairwise\\
         Matching Keyword & +700& & Pairwise\\
         Keyword in token & +70& & Pairwise\\
              Extra Token & & -500& Pairwise\\
Matching token in context & +10& & Pairwise\\
Similar neighbor & +100&-5 & Entity-wide\\
No Similar neighbor in entity & & -5& Entity-wide\\
Matching document & +350&-15 & Entity-wide\\
No Matching documents in entity & & -15& Entity-wide\\
\hline

\end{tabularx}
\label{table:features}
\end{center}
\end{table}

\paragraph{Models.}
Features on the NYT and Wikilinks data sets were manually tuned and the
features for the Rexa data set were trained using sample
rank~\cite{wick2011samplerank} with confidence weighted updates.
We manually tune some of the weights in the NYT corpus to make up for the lack the complete training data.
The models can be graphically represented as the models in Figure~\ref{fig:query_driven_resolved}.

\paragraph{Evaluation metrics.}
Convergence of MCMC algorithms is difficult to measure as describe in a review by Cowles and
Carlin~\cite{cowles1996markov}.
We estimate the convergence progress by calculating the $f1$ score of the query node's entity ($f1_q$).
We create this new measure because we are primarily concerned with the query entity.
Other measures include B$^3$ for entity resolution and several others for general MCMC models~\cite{bagga1998entity, cowles1996markov}.

The query-specific $f1$ score is the harmonic mean of the query-specific recall $R_q$ and query-specific precision $P_q$.
To accurately determine the $P_q$ and $R_q$ of each
query in this experiment we label each correct query node.
Query-specific precision is defined as $P_q = \frac{|\{\textrm{relevant}(\mathcal{M})\} \cap \{\textrm{retrieved}(\mathcal{M})\}|}{|\{\textrm{retrieved}(\mathcal{M})\}|}$
and query-specific recall $ R_q = \frac{|\{\textrm{relevant}(\mathcal{M})\} \cap \{\textrm{retrieved}(\mathcal{M})\}|}{|\{\textrm{relevant}(\mathcal{M})\}|}$.
The $f1$ score for the query node's entity $q$ is defined as: $$ f1_q = 2\frac{{R_q} {P_q}}{R_q + P_q}.$$
The $f1_q$ score is a good indicator of entity and answer quality.
For multi-query experiments we calculate the average $f1_q$ scores for each query node.
The run of each non-parallel algorithm is averaged over 3 to 10 runs.

\subsection{Realtime Query-Driven ER Over NYT}
\label{sec:eval}

In this experiment we show that query-driven entity resolution techniques allow us to obtain
near realtime\footnote{We define {realtime} as only contributing a
small or no time loss when this process a part of an
external execution pipeline such as an information extraction pipeline.}
results on large data sets such as the NYT corpus.

\begin{figure}
  \centering
  \includegraphics[width=0.5\textwidth]{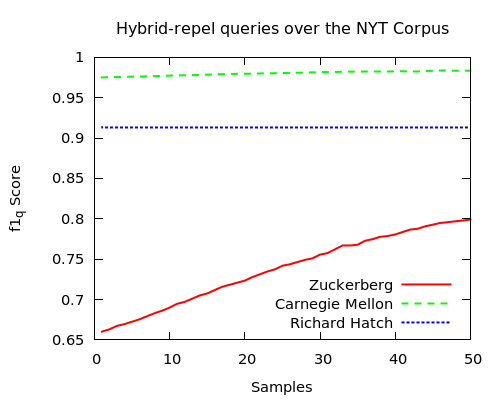}
  \caption{Hybrid-repel performance for the first 50 samples for three queries. Each result is averaged over 6 runs}
  \label{fig:coinfluence1}
\end{figure}

Figure~\ref{fig:coinfluence1} shows the $f1_q$ score of the hybrid ER algorithms
with three single-query ER queries.
The graph shows performance over the first 50 proposals.
For example, the `Zuckerberg' query could be expressed as shown in Algorithm~\ref{alg:selectzuck}.

\begin{algorithm}
\begin{lstlisting}[language=SQL,breaklines=true,frame=none,basicstyle=\normalsize, keywordstyle=\ttfamily, identifierstyle=\ttfamily\bfseries, commentstyle=\color{gray},showstringspaces=false]
SELECT * FROM
Mention m
WHERE coref_map(m.*, entity^p), `zuckerberg', context).
\end{lstlisting}
\caption{Example ER query over the entity `zuckerberg'}
\label{alg:selectzuck}
\end{algorithm}

Recall, a canopy is first generated using an approximate match over the mention set.
We use the repel inference function and all the mentions are initialized in one large entity.
The `Richard Hatch' and `Carnegie Mellon' queries start at an $f1_q$ score of $.92$ and $.97$, respectively.
The `Zuckerberg' query starts above $.65$ and improves to an $f1_q$ score over $.8$.

These experiments show the repel method removing mismatches from the
query entity. The co-influence function is used to quickly identify
the mentions that do not belong in the entity and they are proposed
to be removed. When a hybrid move is proposed, a mention from the
large entity moved from a large entity group to a new, possibly
empty, entity. This method relies on the good repulsion features and
correct weights.

In Table~\ref{tab:realtime1} we show the performance of three queries.
In addition to the query token we add four columns: blocking time in seconds, canopy size, inference time in
seconds and the total compute time.
Total time is the complete time taken by each run, this includes building of the influence data structure and
result writing.
The values in Table~\ref{tab:realtime1} show that fast performance of query-driven ER over a large
database of mentions.

\begin{table}
\centering
  \caption[The performance of the hybrid-repel ER algorithm for queries over the NYT corpus for the first 50 samples.]{The performance of the hybrid-repel ER algorithm for queries over the NYT corpus for the first 50 samples.
Total time includes the time to build the $\bar{\mathcal{I}}$ data structure and result output.
The NYT Corpus contains over 71 million mentions, a large amount for the entity resolution problems.}
  \begin{tabularx}{\textwidth}{XXXXXX}
    \hline
    {Query} & {Blocking} & {Mentions} & {Inference} & {Total time}\\
    \hline
    Zuckerberg & 24.4 s & 103 &  2 s & 37 s\\
    Richard Hatch & 28.3 s & 226 & 18.5 s & 59 s \\
    Carnegie Mellon & 25.9 s & 1302 & 68 s & 124 s \\
    \hline
  \end{tabularx}
  \label{tab:realtime1}
\end{table}

\subsection{Single-query ER}
In this experiment we show a performance comparison between the single-query algorithms summarized
in Sections~\ref{sec:algorithms} and~\ref{sec:queryproportional}.
We run the query-driven algorithms over queries with different selectivity levels and show
the accuracy over time.
Each algorithm uses the attract method, so each mention in the canopy starts in its own entity.

\begin{figure}
    \centering
    \includegraphics[width=0.5\textwidth, clip=true]{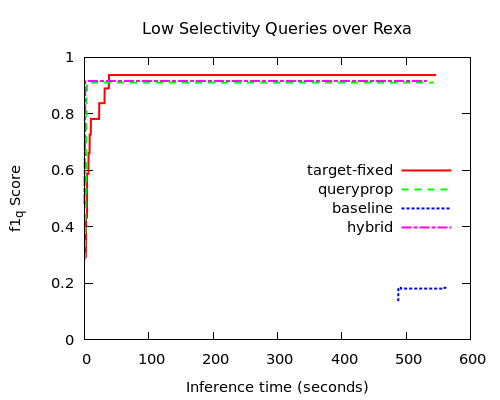}
    \caption{A comparison of single-query algorithms on a query with selectivity of 11}
\label{fig:exp1rexavsmall}
  \end{figure}

Figure~\ref{fig:exp1rexavsmall} shows the run time of all four algorithms on the Rexa data set with
the query `Nemo Semret', an author with a selectivity of 11.
The performance for the baseline entity resolution does not get a correct proposal
until about 500 seconds.
The baseline algorithm takes a long time to accept the first proposal because it is randomly
trying to insert mentions into  an existing entity.
Target-fixed immediately begins to make correct proposals.
Hybrid and query-proportional have the best performance and resolve the entity almost instantaneously.
The hybrid chooses the most likely nodes to merge into the query entity.
As the first couple of proposals are correct merges, hybrid quickly converges
to a high accuracy.
Due to imperfect features, among the 10 averaged runs
a few runs get stuck at local optimum and causing suboptimal results.

\begin{figure}
    \centering
    \includegraphics[width=0.5\textwidth, clip=true]{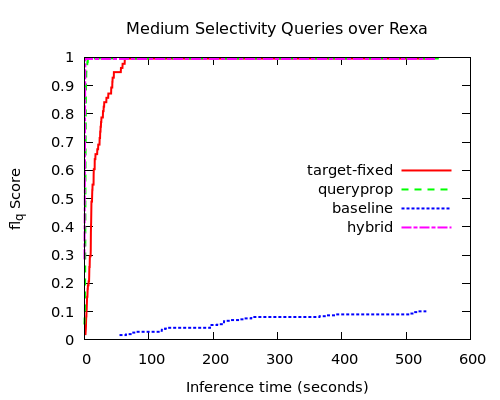}
    \caption{A comparison of single-query algorithms with a query node of selectivity 46}
\label{fig:exp1rexasmall}
  \end{figure}

Figure~\ref{fig:exp1rexasmall} shows the run time of four algorithms for query node id `A. A. Lazar'
with selectivity of 46.
The baseline algorithm progresses the slowest.
The hybrid algorithm quickly reaches a perfect $f1_q$ score.
Query-proportional algorithm lags slightly behind the hybrid
method but still reaches a perfect value.
The target-fixed algorithm gradually increases to a perfect $f1_q$ score about 60 seconds
after hybrid and query-proportional.

  \begin{figure}
    \centering
    \includegraphics[width=0.5\textwidth, clip=true]{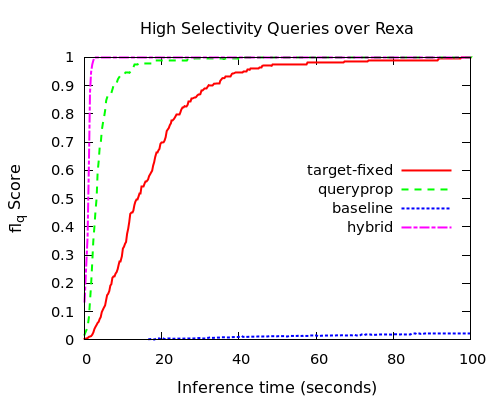}
    \caption{A comparison of selection-driven algorithms with a query node of selectivity 130}
\label{fig:exp1rexa}
  \end{figure}

Figure~\ref{fig:exp1rexa} shows the run time of four algorithms with
a query `Michael Jordan' of selectivity 130. The baseline slowly
increases over the 100 seconds. The hybrid algorithm again quickly
achieves a perfect $f1_q$ score followed by query-proportional and
then target-fixed. The time gap between each of the algorithms
increases with the increase in selectivity, hybrid achieves the best
performance.

 \begin{figure}
    \centering
    \includegraphics[width=0.5\textwidth, clip=true]{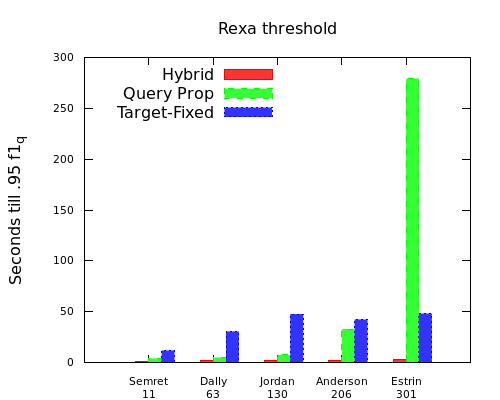}
    \caption{The time until an $f1_q$ score of 0.95 for five queries of increasing selectivities; averaged over three runs}
    \label{fig:rexaselectivity}
  \end{figure}

We look deeper at how selectivity affects the rate of convergence.
In Figure~\ref{fig:rexaselectivity} we show the time it takes for
each algorithm to reach an $f1_q$ score of $0.95$ over increasing 
selectivity. We choose five query nodes of increasing selectivity
but with the same canopy sizes. The hybrid algorithm runtime
increased with the increase in selectivity but only slightly steeper
than constant. Target-fixed increased for the first three queries
but did not last more than 50 seconds. Query-proportional has only a
slight increase in time till convergence for the first three
queries. The highest two selectivity queries are expensive for
query-proportional and we observe an exponential increase in
runtime. These results are consistent with the exponentially large
increase in the number of random comparisons needed to find a match
for a query entity. The query-proportional algorithm does not focus
on the query entity as aggressively as target-fixed and hybrid
algorithms.
Recall that the target-fixed and the hybrid algorithm focus on moving correct nodes
into the query entity. Query-proportional
selects candidate nodes using the influence function but it does not fix the target entity.
With the target entity not fixed, the chance of correct node for the query entity decrease exponentially.
This shows that selectivity of nodes affects the runtime performance of each algorithm.
When performing join-driven ER it is important to take the relative selectivity
of nodes into account for choosing best scheduling algorithms.

\subsection{Multi-query ER}
\begin{figure}
  \centering
  \includegraphics[width=0.5\textwidth]{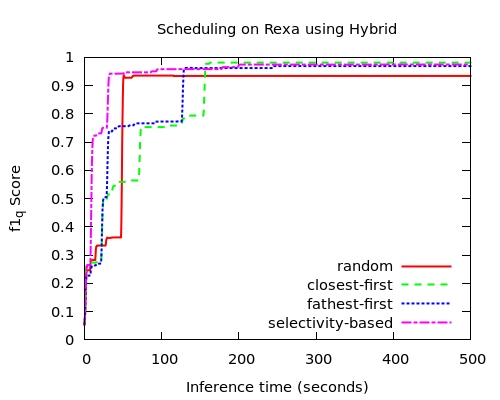}
  \caption{The progress of the hybrid algorithm across for multiple query nodes using difference scheduling algorithms. Each result is averaged over three runs}
  \label{fig:rexaschedule}
\end{figure}

In this experiment we study performance of our different scheduling
algorithms for join-driven ER queries. We choose ten query nodes of
different selectivity and run the join queries scheduling algorithms
described in Section~\ref{sec:jointquery}. Consider a table like the
{People} table in Section~\ref{sec:problemstatement} with
selectivity
 \{130, 63, 68, 7, 12, 12, 301, 11, 46\}.
The four algorithms, random, closest-first, farthest-first and
selectivity-based are shown in Figure~\ref{fig:rexaschedule}.
The selectivity-based method out performs the
other three algorithms in terms of convergence rate.
The jumps in accuracy on the graph correspond to the scheduling algorithms choosing new query nodes and accepting
new proposals.
It has a high jump when it starts sampling the seventh, and highest selectivity nodes.
The farthest-first algorithm rises the slowest out of the scheduling algorithms because it tries to
stop sampling the high performing query entity and makes proposals for the slowest growing.
Selectivity-based method performs well early because the high selectivity queries are  sampled first.
The high selectivity query makes up a large proportion of the total $f1_q$ score.
The large jump in the random method is when it reaches the node with selectivity 301.
Notice, closest-first reaches its peak $f1_q$ score the fastest because it tries to get the most out
of every query node.

\subsection{Context Levels}
\begin{figure}
  \centering
  \includegraphics[width=0.5\textwidth]{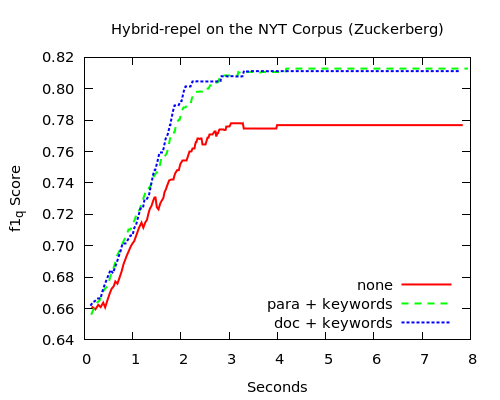}
  \caption{The performance of {zuckerberg} query with difference levels of context. Each result is averaged over 6 runs}
  \label{fig:coinfluence2.png}
\end{figure}

In this experiment we aim to discover how different levels of context specified at query time can improve
convergence time and overall accuracy.
We take the {zuckerberg} query and the hybrid-repel algorithm and ran ER
three times over three levels of context.
Each mention in the graph contains a `paragraph' level of context
and we only alter the context of the query node.
The `none' context only activates token specific features, any context features involving the
query node are zeroed out.
The `paragraph' level context is the default context from the NYT corpus and the `document' level context extends
context to the entire news article.
Additionally, we add specific keywords from Mark Zuckerberg's DBpedia page to
the `document' and `paragraph' context levels.
We show the performance using the repel method in Figure~\ref{fig:coinfluence2.png}.

Adding specific keywords that activate the keyword features are the most effective methods for increasing
the accuracy of query-driven ER\@.
Query-driven methods allow a user to observe the results and add or remove keywords specific
queries to improve the accuracy.
This type of iterative improvement workflow is not feasible with batch methods.

\subsection{Parallel Hybrid ER}
\label{sec:parallelexperiments}
\begin{figure}
  \centering
  \includegraphics[width=0.5\textwidth]{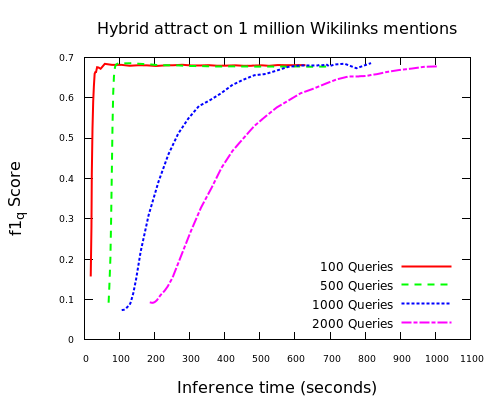}
  \caption{Hybrid-attract algorithm with random queries run over the Wikilinks corpus. Each plot starts after the Vose structures are constructed}
  \label{fig:crossover.png}
\end{figure}

In this experiment has two objectives, first how does the hybrid algorithm perform
in a canopy size of 1 million queries and what is the effect of increasing the number of queries nodes.
In Figure~\ref{fig:crossover.png} the Hybrid algorithm is able to resolve entities in a short amount
of time. 
The creation time of the Vose structure is about linear in the number of queries.
The trend in the graph is that as the ratio of queries to entities increases the performance benefit
of the hybrid-attract method decreases.
With more query nodes the construction time increases and the benefits of the algorithm decrease
and become no better than the baseline method.

\paragraph{Experiment Summary}
Each of the query-driven methods outperform the baseline methods in terms of
runtime while not losing out on accuracy.
Across different data set sizes hybrid algorithms have the most consistent performance.
If a system has a quality blocking function then it is better to use the co-influence entity resolution method.
With multiple query nodes, selectivity-based is the most consistent performing algorithm.
More accurate estimation of MCMC convergence performance could allow the
dynamic scheduling algorithms closest-first and farthest-first to achieve higher accuracy.
The more contextual information that can be added to query nodes at query time causes higher accuracy of
the entity resolution algorithms.
Parallel query-driven sampling is an effective way to get speed up in an ER data set when
the ratio of mentions to entities is low.

\section{Query-Driven Entity Resolution Related Work}
\label{sec:qderrelatedwork}

This chapter is related to work in several areas.
In this section we describe a selection of the literature that we found most relevant to different parts of the Query-Driven ER task.

\paragraph{Entity Resolution}
The state-of-the-art method for entity resolution employs collective classification.
Instead of purely pairwise decisions, collective classification methods consider group relationships
when making clustering determinations.
In a recent tutorial~\cite{ervldb12tutorial}, collected classification methods were grouped into three categories:
non-probabilistic~\cite{Dong:2005:RRC:1066157.1066168,Kalashnikov:2006:DDC:1138394.1138401, Bhattacharya:2007:CER:1217299.1217304},
  probabilistic~\cite{li2004identification, pasula2002identity, getoor2006latent, mccallum2004nips, singla2006entity, DBLP:conf/uai/BrochelerMG10} and
  hybrid approaches~\cite{shen2005constraint, arasu2009large}.
  A relevant challenge proposed for entity resolution research by the tutorial is how to
  efficiently perform entity resolution when a query is involved.
This chapter seeks to address this issue.

Entity resolution is generally an expensive, offline batch process.
Bhattacharya and Getoor proposed a method for query-time entity resolution~\cite{Bhattacharya:2006:QER:1150402.1150463}.
This method performs inference by starting with a query node and performing `expand and resolve' to resolve entities through
resolution of attributes and expansion of hyper-edges.
Unfortunately, hyper-edges between records are not always explicit in data sets.
This chapter does not assume the presence of any link in the corpus, each entity or mentions are independently defined, which is the case for most applications.

A recent paper by Altwaijry, Kalashnikov and Mehrotra~\cite{altwaijry2013query} has a similar motivation
of using SQL queries to drive entity resolution.
That work focuses using predicates in the query to drive computation while this work uses example
queries to drive computation.
Both techniques are complementary and combining the two by updating the {edge-picking}
policy described in their paper using our approach makes for interesting method of optimizing the entity resolution process.

The term {query-driven} appears in this chapter and has appeared in others 
across literature with different meanings~\cite{Grant:2012:MQS:2396761.2398746}.
Our definition of a query node is an example item, mention, from a data set. 
A query in Altwaijry et al.~\cite{altwaijry2013query} are the predicates in an SQL statement. 
Query-driven in Grant et al.~\cite{Grant:2012:MQS:2396761.2398746} is the SQL queries used to drive analytics.

It is becoming increasingly normal to work with data sets of extremely large size, in response researchers
have studied streaming and distributed processing.
Rao, McNamee and Dredze describes an approach for streaming entity resolution~\cite{rao2010streaming}.
This approach is fast and approximates entries in an LRU queue of clustered entity chains.
We apply these techniques to static data set and do not yet handle streams of data.
Singh, Subramanya, Pereira, and McCallum propose a technique for ER where entities are resolved in
parallel blocks and then redistributed and resolved again in new blocks~\cite{singh2011large}.
This parallel distribution method makes large-scale entity resolution tractable.
In this chapter, we perform analysis on a similar scale data set but we show that great performance gains can be
achieved when a query is specified.

\paragraph{Query specific sampling}
Recently, several researchers have explored the idea of focusing sampling of graphical models
to speed up inference.
Below we discuss the three approaches that use sampling to speed up ER over graphical models.

Query-Aware MCMC~\cite{qam} found that when performing a query over a graphical model the cost of not
sampling a node is exactly the nodes influence on the query node.
This enables us to ignore some nodes that have low influence over the query node and incur a small amount of error.
This influence score can be calculated as the mutual information between two nodes.
The authors compare estimation techniques of the intractable mutual information
score, this is called the {influence trail score}.
Because ER has a fixed pairwise model, we can use the theory from this work and specialized data structures
to gain performance when query-driven sampling.

Type-based MCMC is a method of sampling groups of nodes of with the same attribute to
increase the progress towards convergence~\cite{Liang:2010:TM:1857999.1858081}.
This approach works well when feature sets can be tractably counted and grouped.
If query nodes are introduced it is not clear how one may focuses type-based sampling.

Other researchers have explored using belief propagation with
queries to approximate marginal of factor graphs~\cite{Chechetka+Guestrin:aistats10qsbp}.
However, the entity resolution graph is cyclic and highly connected. MCMC scales with
large real world models better than loopy belief propagation~\cite{qam}.


%

%

\section{Query-Driven Entity Resolution Summary}
\label{sec:conclusion}

In this chapter, I propose new approaches for accelerating large-scale entity resolution 
in the common case that the user is interested in one or a watch list of entities.
These techniques can be integrated into existing data processing pipelines or used as a tool for exploratory data analysis.
We showed three single-node ER algorithms and three scheduling algorithms for multi-query 
ER and show experimentally how their runtime performance is several orders of magnitude better
than the baseline.

\balance

\bibliographystyle{abbrv}


  \bibliography{citation}

\begin{thebibliography}{10}

\bibitem{altwaijry2013query}
H.~Altwaijry, D.~V. Kalashnikov, and S.~Mehrotra.
\newblock Query-driven approach to entity resolution.
\newblock {\em Proceedings of the VLDB Endowment}, 6(14), 2013.

\bibitem{arasu2009large}
A.~Arasu, R.~Christopher, and D.~Suciu.
\newblock Large-scale deduplication with constraints using dedupalog.
\newblock In {\em International Conference on Data Engineering}, pages
  952--963. IEEE, 2009.

\bibitem{Arumugam:2010:DSD:1807167.1807224}
S.~Arumugam, A.~Dobra, C.~M. Jermaine, N.~Pansare, and L.~Perez.
\newblock The datapath system: A data-centric analytic processing engine for
  large data warehouses.
\newblock In {\em Proc. of the 2010 ACM SIGMOD}, pages 519--530, NY, USA, 2010.
  ACM.

\bibitem{bagga1998entity}
A.~Bagga and B.~Baldwin.
\newblock Entity-based cross-document coreferencing using the vector space
  model.
\newblock In {\em 17th ACL}, pages 79--85. ACL, 1998.

\bibitem{Bhattacharya:2007:CER:1217299.1217304}
I.~Bhattacharya and L.~Getoor.
\newblock Collective entity resolution in relational data.
\newblock {\em ACM Trans. KDD}, 1(1), Mar. 2007.

\bibitem{Bhattacharya:2006:QER:1150402.1150463}
I.~Bhattacharya, L.~Getoor, and L.~Licamele.
\newblock Query-time entity resolution.
\newblock In {\em Proc.12th ACM SIGKDD}, KDD '06, pages 529--534, NY, USA,
  2006.

\bibitem{nltk09}
S.~Bird, E.~Loper, and E.~Klein.
\newblock Natural language processing with python.
\newblock In {\em Natural Language Processing with Python}. O'Reilly Media Inc,
  2009.

\bibitem{DBLP:conf/uai/BrochelerMG10}
M.~Br{\"o}cheler, L.~Mihalkova, and L.~Getoor.
\newblock Probabilistic similarity logic.
\newblock In P.~Gr{\"u}nwald and P.~Spirtes, editors, {\em UAI}, pages 73--82.
  AUAI Press, 2010.

\bibitem{Chechetka+Guestrin:aistats10qsbp}
A.~Chechetka and C.~Guestrin.
\newblock Focused belief propagation for query-specific inference.
\newblock In {\em AISTATS}, May 2010.

\bibitem{cowles1996markov}
M.~Cowles and B.~Carlin.
\newblock Markov chain monte carlo convergence diagnostics: a comparative
  review.
\newblock {\em Journal of AmStat}, 91(434):883--904, 1996.

\bibitem{DasSarma:2012:ABM:2396761.2398403}
A.~Das~Sarma, A.~Jain, A.~Machanavajjhala, and P.~Bohannon.
\newblock An automatic blocking mechanism for large-scale de-duplication tasks.
\newblock In {\em Proceedings of the 21st ACM CIKM}, pages 1055--1064. ACM,
  2012.

\bibitem{Dong:2005:RRC:1066157.1066168}
X.~Dong, A.~Halevy, and J.~Madhavan.
\newblock Reference reconciliation in complex information spaces.
\newblock In {\em Proceedings of the 2005 ACM SIGMOD international conference
  on Management of data}, SIGMOD '05, pages 85--96, New York, NY, USA, 2005.
  ACM.

\bibitem{getoor2006latent}
I.~Getoor.
\newblock A latent dirichlet model for unsupervised entity resolution.
\newblock In {\em Proceedings of the 6th SIAM International Conference on Data
  Mining}, volume 124, page~47. Society for Industrial Mathematics, 2006.

\bibitem{ervldb12tutorial}
L.~Getoor and A.~Machanavajjhala.
\newblock Entity resolution: Theory, practice \& open challenges.
\newblock In {\em Proceedings of the 38rd VLDB}, VLDB '12. VLDB Endowment,
  2012.

\bibitem{graff2007ldc2007t07}
D.~Graff.
\newblock Ldc2007t07: English gigaword corpus, 2007.

\bibitem{Grant:2012:MQS:2396761.2398746}
C.~E. Grant, J.-d. Gumbs, K.~Li, D.~Z. Wang, and G.~Chitouras.
\newblock Madden: query-driven statistical text analytics.
\newblock In {\em Proceedings of the 21st ACM international conference on
  Information and knowledge management}, CIKM '12, pages 2740--2742, New York,
  NY, USA, 2012. ACM.

\bibitem{gravano2001using}
L.~Gravano, P.~Ipeirotis, H.~Jagadish, N.~Koudas, S.~Muthukrishnan,
  L.~Pietarinen, and D.~Srivastava.
\newblock Using q-grams in a dbms for approximate string processing.
\newblock {\em IEEE Data Engineering Bulletin}, 24(4):28--34, 2001.

\bibitem{hellerstein2012madlib}
J.~M. Hellerstein, C.~R{\'e}, F.~Schoppmann, D.~Z. Wang, E.~Fratkin,
  A.~Gorajek, K.~S. Ng, C.~Welton, X.~Feng, K.~Li, and A.~Kumar.
\newblock The madlib analytics library: or mad skills, the sql.
\newblock {\em Proceedings of the VLDB Endowment}, 5(12):1700--1711, Aug. 2012.

\bibitem{Jampani:2008:MMC:1376616.1376686}
R.~Jampani, F.~Xu, M.~Wu, L.~L. Perez, C.~Jermaine, and P.~J. Haas.
\newblock Mcdb: A monte carlo approach to managing uncertain data.
\newblock In {\em Proc. of the 2008 ACM SIGMOD}, pages 687--700, NY, USA, 2008.
  ACM.

\bibitem{Kalashnikov:2006:DDC:1138394.1138401}
D.~V. Kalashnikov and S.~Mehrotra.
\newblock Domain-independent data cleaning via analysis of entity-relationship
  graph.
\newblock {\em ACM Trans. Database Syst.}, 31(2):716--767, June 2006.

\bibitem{koller2009probabilistic}
D.~Koller and N.~Friedman.
\newblock {\em Probabilistic graphical models: principles and techniques}.
\newblock MIT press, 2009.

\bibitem{li2013gptext}
K.~Li, C.~Grant, D.~Z. Wang, S.~Khatri, and G.~Chitouras.
\newblock Gptext: Greenplum parallel statistical text analysis framework.
\newblock In {\em Proceedings of the Second Workshop on Data Analytics in the
  Cloud}, pages 31--35. ACM, 2013.

\bibitem{li2004identification}
X.~Li, P.~Morie, and D.~Roth.
\newblock Identification and tracing of ambiguous names: Discriminative and
  generative approaches.
\newblock In {\em Proceedings of the National Conference on Artificial
  Intelligence}, pages 419--424. Menlo Park, CA; Cambridge, MA; London; AAAI
  Press; MIT Press; 1999, 2004.

\bibitem{Liang:2010:TM:1857999.1858081}
P.~Liang, M.~I. Jordan, and D.~Klein.
\newblock Type-based mcmc.
\newblock In {\em Human Language Technologies: The 2010 NAACL}, HLT '10, pages
  573--581, Stroudsburg, PA, USA, 2010. ACL.

\bibitem{dataspace}
J.~Madhavan, S.~Jeffery, S.~Cohen, X.~Dong, D.~Ko, C.~Yu, and A.~Halevy.
\newblock Web-scale data integration: You can only afford to pay as you go.
\newblock In {\em Proceedings of CIDR}, pages 342--350, 2007.

\bibitem{mccallum2000efficient}
A.~McCallum, K.~Nigam, and L.~Ungar.
\newblock Efficient clustering of high-dimensional data sets with application
  to reference matching.
\newblock In {\em Proc. of the 6th SIGKDD}, pages 169--178, 2000.

\bibitem{mccallum09:factorie}
A.~McCallum, K.~Schultz, and S.~Singh.
\newblock {FACTORIE}: Probabilistic programming via imperatively defined factor
  graphs.
\newblock In {\em NIPS}, pages 1426--1427, 2009.

\bibitem{mccallum2004nips}
A.~Mccallum and B.~Wellner.
\newblock {Conditional Models of Identity Uncertainty with Application to Noun
  Coreference}.
\newblock In {\em NIPS}, 2004.

\bibitem{pasula2002identity}
H.~Pasula, B.~Marthi, B.~Milch, S.~J. Russell, and I.~Shpitser.
\newblock {Identity Uncertainty and Citation Matching}.
\newblock In {\em Neural Information Processing Systems}, pages 1401--1408,
  2002.

\bibitem{rao2010streaming}
D.~Rao, P.~McNamee, and M.~Dredze.
\newblock Streaming cross document entity coreference resolution.
\newblock In {\em Proceedings of the 23rd International Conference on
  Computational Linguistics: Posters}, pages 1050--1058. Association for
  Computational Linguistics, 2010.

\bibitem{shen2005constraint}
W.~Shen, X.~Li, and A.~Doan.
\newblock Constraint-based entity matching.
\newblock In {\em Proceedings of the National Conference on Artificial
  Intelligence}, volume~20, page 862. Menlo Park, CA; Cambridge, MA; London;
  AAAI Press; MIT Press; 1999, 2005.

\bibitem{5767835}
L.~Shu, A.~Chen, M.~Xiong, and W.~Meng.
\newblock Efficient spectral neighborhood blocking for entity resolution.
\newblock In {\em 2011 IEEE 27th ICDE}, pages 1067 --1078, april 2011.

\bibitem{singh2011large}
S.~Singh, A.~Subramanya, F.~Pereira, and A.~McCallum.
\newblock Large-scale cross-document coreference using distributed inference
  and hierarchical models.
\newblock In {\em Proceedings of the 49th Annual Meeting of the Association for
  Computational Linguistics: Human Language Technologies-Volume 1}, pages
  793--803. Association for Computational Linguistics, 2011.

\bibitem{singh12:wiki-links}
S.~Singh, A.~Subramanya, F.~Pereira, and A.~McCallum.
\newblock Wikilinks: A large-scale cross-document coreference corpus labeled
  via links to {Wikipedia}.
\newblock Technical Report UM-CS-2012-015, 2012.

\bibitem{singla2006entity}
P.~Singla and P.~Domingos.
\newblock {Entity Resolution with Markov Logic}.
\newblock In {\em IEEE International Conference on Data Mining}, pages
  572--582. IEEE, 2006.

\bibitem{Vose:1991:LAG:126262.126280}
M.~D. Vose.
\newblock A linear algorithm for generating random numbers with a given
  distribution.
\newblock {\em IEEE Trans. Softw. Eng.}, 17(9):972--975, Sept. 1991.

\bibitem{Whang:2009:ERI:1559845.1559870}
S.~E. Whang, D.~Menestrina, G.~Koutrika, M.~Theobald, and H.~Garcia-Molina.
\newblock Entity resolution with iterative blocking.
\newblock In {\em 2009 ACM SIGMOD}, pages 219--232. ACM, 2009.

\bibitem{Wick:2010:SPD:1920841.1920942}
M.~Wick, A.~McCallum, and G.~Miklau.
\newblock Scalable probabilistic databases with factor graphs and mcmc.
\newblock {\em Proc. VLDB Endow.}, 3(1-2):794--804, Sept. 2010.

\bibitem{Wick:2012:DHM:2390524.2390578}
M.~Wick, S.~Singh, and A.~McCallum.
\newblock A discriminative hierarchical model for fast coreference at large
  scale.
\newblock In {\em Proceedings of the 50th ACL}, ACL '12, pages 379--388, 2012.

\bibitem{qam}
M.~L. Wick and A.~McCallum.
\newblock Query-aware mcmc.
\newblock In J.~Shawe-Taylor, R.~Zemel, P.~Bartlett, F.~Pereira, and
  K.~Weinberger, editors, {\em Advances in NIPS 24}, pages 2564--2572, 2011.

\bibitem{wick2011samplerank}
M.~L. Wick, K.~Rohanimanesh, K.~Bellare, A.~Culotta, A.~McCallum, and
  A.~McCallum.
\newblock Sample rank: Training factor graphs with atomic gradients.
\newblock In {\em ICML}, pages 777--784, 2011.

\end{thebibliography}




\end{document}